\documentclass[useAMS,usenatbib]{mn2e}

\pdfoutput=1

\usepackage{graphicx,natbib,aas_macros,amssymb,lscape,epsfig,array,multirow,epsfig,rotating,lscape,enumerate}

\def\tablenotemark#1{\rlap{$^{#1}$}}
\def\tablenotetext#1#2{
\@temptokena={\vspace{.5ex}{\noindent\llap{$^{#1}$}#2}\par}
\@temptokenb=\expandafter{\tblnote@list}
\xdef\tblnote@list{\the\@temptokenb\the\@temptokena}}

\title[Tides and Tidal Engulfment in Post-Main Sequence Binaries]{Tides and Tidal Engulfment in Post-Main Sequence Binaries: Period Gaps for Planets and Brown Dwarfs Around White Dwarfs}

\author[J.~Nordhaus et al.]{J. Nordhaus$^1$\thanks{E-mail:
nordhaus@astro.princeton.edu}, D. S. Spiegel$^{1,2}$, L. Ibgui$^{1,3}$, J. Goodman$^1$, A. Burrows$^{1,2}$\\
$^1$Department of Astrophysical Sciences, Princeton University, Princeton, NJ 08544, U.S.A.\\
$^2$Kavli Institute for Theoretical Physics, UCSB, Santa Barbara, CA 93106, U.S.A.\\
 $^3$ LERMA, Observatoire de Paris, CNRS et UPMC, 5 place J.
Janssen, 92195 Meudon, France}

\begin{document}
\date{Submitted 25 February 2010}
\pubyear{2010}
\maketitle
\label{firstpage}

\begin{abstract}
The presence of a close, low-mass companion is thought to play a
substantial and perhaps necessary role in shaping post-Asymptotic
Giant Branch and Planetary Nebula outflows.  During post-main-sequence
evolution, radial expansion of the primary star, accompanied by
intense winds, can significantly alter the binary orbit via tidal
dissipation and mass loss.  To investigate this, we couple stellar
evolution models (from the zero-age main-sequence through the end of
the post-main sequence) to a tidal evolution code.  The binary's fate
is determined by the initial masses of the primary and the companion,
the initial orbit (taken to be circular), and the Reimers mass-loss
parameter.  For a range of these parameters, we determine whether the
orbit expands due to mass loss or decays due to tidal torques.  Where
a common envelope (CE) phase ensues, we estimate the final orbital
separation based on the energy required to unbind the envelope.  These
calculations predict period gaps for planetary and brown dwarf
companions to white dwarfs.  The upper end of the gap is the shortest
period at which a CE phase is avoided.  The lower end is the longest
period at which companions survive their CE phase.  For binary systems
with 1~$M_\odot$ progenitors, we predict no Jupiter-mass companions
with periods $\lesssim$270 days.  Once engulfed, Jupiter-mass companions do not survive a CE phase.  For binary systems consisting of a 1~$M_\odot$ progenitor with a companion 10 times the mass of Jupiter,
we predict a period gap between $\sim$0.1 and $\sim$380 days.  These results are
consistent with both the detection of a $\sim$50 $M_{\rm J}$ brown dwarf in a $\sim$0.003 AU ($\sim$0.08 day) orbit around the white dwarf WD 0137-349 and the tentative detection of a $\sim$2~$M_{\rm J}$
planet in a $\gtrsim$2.7~AU ($\gtrsim$4~year) orbit around the white dwarf GD66.
\end{abstract}


\section{Introduction}
\label{sec:intro}
For low- and intermediate-mass stars (initially
$\lesssim$8~M$_\odot$), post-main sequence (post-MS) evolution is
characterized by expansion via giant phases accompanied by the onset
of mass-loss.  In particular, during the Asymptotic Giant Branch phase
(AGB), dust-driven winds expel the stellar envelope as the star begins
its transition to a white dwarf (WD).  However, before formation of
the WD remnant, the spherical outflows observed during the AGB phase
undergo a dramatic transition to the highly asymmetric and often
bipolar geometries seen in the post-AGB and planetary nebula phases
(PN; \citealt{Sahai:1998ee}).  This transition is often accompanied by
high-speed, collimated outflows.  For recent reviews see
\citet{Balick:2002yf}, \citet{van-Winckel:2003pi},
\citet{de-Marco:2009vl}.

A central hypothesis to explain shaping in post-AGB/PNe is that a
close companion is necessary to power and shape bipolarity.  This is
supported by observations of excess momenta in almost all post-AGB
outflows relative to what isotropic radiation pressure can provide
\citep{Bujarrabal:2001bs}.  Additionally, close companions have been
detected in or around giants, bipolar post-AGB and PNe systems and in
some cases seem to be responsible for outflow shaping
\citep{Silvotti:2007fk, De-Marco:2008nx, Sato:2008qy, Sato:2008uq,
  Witt:2009wd, Miszalski:2009oq, Miszalski:2009eu, Niedzielski:2009lr,
  Chesneau:2009lr}.  In particular, the recent detection of a white dwarf with an orbiting $\sim$50 $M_{\rm J}$ brown dwarf in a $\sim$2 hour orbit demonstrates that low-mass companions can survive a common envelope phase (CEP) \citep{Maxted:2006fj}.  The detection of a planetary companion ($M{\rm sin}i=3.2$ $M_{\rm J}$) around the extreme horizontal branch star V391 Pegasi in a $\sim$1.7 AU orbit ($\sim$3.2 year period; \citealt{Silvotti:2007fk}) and the tentative detection of a $\sim$2~$M_{\rm J}$ planet in a $\gtrsim$2.7~AU orbit ($\gtrsim$4~year period; \citealt{Mullally:2008fk,Mullally:2009uq}) around the white dwarf GD66 provide motivation for this work.

Indirect observational evidence for binarity comes from maser
observations that suggest magnetic jet collimation in AGB and young
post-AGB stars \citep{Vlemmings:2006vl,Sabin:2007zp,Vlemmings:2008ad}.
Such collimation supports the binary hypothesis because it is
difficult for single AGB stars to generate the large field strengths
necessary to power the outflows without an additional source of
angular momentum \citep{Nordhaus:2007il,Nordhaus:2008fe}.  If a close
companion is present, strong interactions can transfer energy and
angular momentum from the companion to the primary.  In particular, if
the companion is engulfed in a common envelope (CE), rapid in-spiral
can cause significant differential rotation of the envelope
\citep{Nordhaus:2006oq, Nordhaus:2008lr}.  Coupled with a strong
convective envelope, large-scale magnetic fields are amplified and are
sufficient to unbind the envelope and power the outflow
\citep{Nordhaus:2007il}.

\cite{Moe:2006fc} overpredicted the Galactic PN population by a factor
of $\sim$6 for PN with radii $\lesssim$0.9~pc (discrepant at the 3
$\sigma$ level; \citealt{Jacoby:1980kc,Peimbert:1990ta}).  This
discrepancy could be alleviated if only a fraction of stars become
PNe.  The authors argue that perhaps interacting binaries or, in
particular, common envelope systems may be responsible for producing
the majority of Galactic PNe \citep{De-Marco:2005xw}.  However, to
determine the validity of a hypothetical CE/PN connection, it is
important to understand which binary systems will undergo a CE phase
during their evolution.  The fact that low-mass stellar companions can
survive CE evolution motivates further study independent of a
potential connection to PNe \citep{Maxted:2006fj}.

Various aspects of tides on the evolution of post-MS binaries have
previously been investigated \citep{Rasio:1996yq, Carlberg:2009uq,
  Villaver:2009qy}.  Recently, \citet{Carlberg:2009uq} aimed to
understand fast rotation in field RGB stars, while
\citet{Villaver:2009qy} focused on trying to predict the distribution
of planets around evolved stars.  \cite{Carlberg:2009uq} calculated
orbital evolution scenarios but neglected mass-loss effects.
\cite{Villaver:2009qy} included stellar mass-loss in their model and
modeled a range of progenitor masses (from 1 to 5~$M_\odot$), but did not examine how results depend on the prescriptions for either tidal interactions or stellar mass loss.  They considered a
wide range of possible astrophysical effects, including frictional and
gravitational drag, wind accretion, and atmospheric evaporation.  We neglect these effects in our study; most are negligible compared to tidal torques and mass loss from the primary.  The exception is perhaps  evaporation, which has been claimed to destroy planets of mass $<15 M_{\rm J}$ \citep{Villaver:2007lr,Livio:1984fk}.  This is similar to the minimum mass required to unbind the stellar envelope (Section 6).  Our work
builds upon both previous results in that we follow the evolution from
the zero-age main sequence (ZAMS) through the entire post-MS, and we
consider a variety of prescriptions for tidal torque and for stellar
mass-loss rates.

\begin{figure}
\begin{center}
\includegraphics[width=8.5cm,angle=0,clip=true]{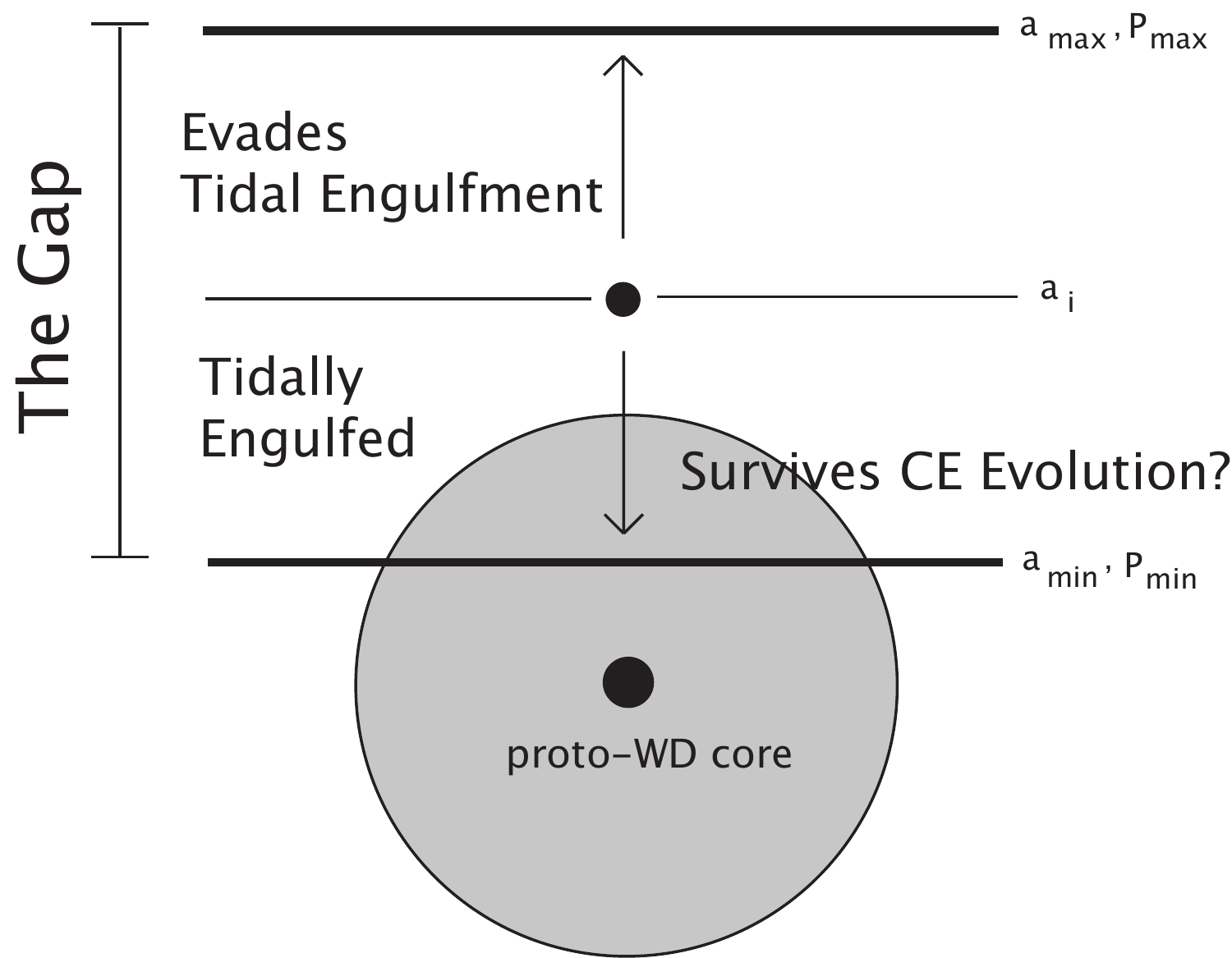}
\caption{The period gap for low-mass companions around white dwarfs.
  The orbit of a companion located initially at $a_i$ decays and
  plunges into the giant star.  Depending on the mass of the companion
  and stellar structure at plunge time, the companion may or may not
  survive the CE phase.  Companions slightly exterior to $a_i$ avoid
  engulfment and never enter a CE; their orbits expand due to
  mass-loss.  The gap is set by the final maximum semimajor axis which
  survives CE evolution ($a_{\rm min}$) and the final minimum
  semimajor axis which avoids tidal engulfment ($a_{\rm max}$).
\label{thegap}}
\end{center}
\end{figure}

In this paper, we investigate how low-mass companions
($\lesssim$0.1~$M_\odot$) become immersed in a common envelope.  By
coupling stellar evolution models to tidal evolution equations, we can
determine when (i) mass-loss dominates and leads to orbital expansion
or (ii) dissipation of tidal energy dominates and leads to orbital
reduction.  In particular, for a given orbital separation, we can
determine which binary systems incur a CE by plunging into their host
stars modulo the uncertainties in the theories of tidal dissipation and stellar mass-loss.  In
\S\ref{sec:2Btides}, we describe our equations and discuss a possible
tidal dependence on period.  In \S\ref{sec:PCmodels}, we present our
stellar evolution models and focus on low-mass companions (planets,
brown dwarfs and low-mass main sequence stars).  In
\S\ref{sec:TResults}, we present our results and discuss the
implications of two commonly used tidal prescriptions.  For systems
that incur a CE, we use an energetics argument to determine the
maximum radius at which a given companion can survive the interaction
by successfully ejecting the envelope.  Furthermore, we determine the
minimum separation that evades a CE phase.  By explicitly following
the orbital evolution, we calculate the final separation of the binary
system.  For a given binary, this produces a gap inside of which we
expect the absence of companions.  From the set of all gaps, we can
determine a minimum period gap for planetary companions around white
dwarfs (see Fig.~\ref{thegap}).  The prediction of a gap should be of
interest to recent searches for both substellar and, in particular,
planetary companions to white dwarfs in addition to future
observations \citep{Mullally:2008fk, Mullally:2009uq, Kilic:2010qy,
  Qian:2010uq}.  We discuss these results in \S\ref{sec:CEevolve} and
\S\ref{sec:PGaps}.  In \S\ref{sec:WD_detect}, we discuss the
possibility of detecting WD+companion transits and applications to
{\it GALEX} data.  We conclude in \S\ref{sec:conc}.

\section{Two-Body Tides}
\label{sec:2Btides}
We employ a two-body gravitational and tidal interaction model that
consistently couples the evolution of the mass and the radius of the
primary star with the orbit of the companion.  It includes the tides
raised on the companion (by the star) and the tides raised on the star
(by the companion).

The evolution of the semimajor axis ($a$) is due to two
contributions: the tidal interaction between the primary and the
companion and the mass loss of the primary
\begin{eqnarray}
\frac{da}{dt} &=& \left( \frac{da}{dt} \right)_{\rm tides} + \left( \frac{da}{dt} \right)_{\rm mass~loss}.\label{eq:da_tot}
\end{eqnarray}
We employ the equations of \cite{Ferraz-Mello:2008dk}, which use the $Q$-formalism of \cite{Goldreich:1963nr} for the tidal
evolution of the orbital eccentricity, $e$ and the semimajor axis,
$a$.  This approach has been used extensively to model the inflated
radii and orbits of transiting extra-solar giant planets (EGP;
\citealt{Kaula:1968fk, Ferraz-Mello:2008dk, Jackson:2008kl,
  Ibgui:2009jk, Miller:2009lr, Ibgui:2009fk,Jackson:2009bq}).

Our formalism assumes that the primary star and companion are both
spherical and that their rotational and orbital angular momenta are
aligned.  Furthermore, we assume the companion spin is synchronized
with its orbital period and that the mean orbital motion is greater
than the stellar rotation rate.  This is reasonable as typical
equatorial surface velocities are $\sim$$1-3$~km~s$^{-1}$ for low-mass
giants implying that, for companions that are close enough for tides
to matter in the post-main-sequence evolution, the angular velocity of
the orbit is much greater than the angular velocity of the primary
star \citep{Gray:1989qa, Massarotti:2008rw}.  Under these assumptions,
we have that
\begin{eqnarray}
\frac{1}{a}\frac{da}{dt} & = & -\frac{1}{a^{13/2}}\left[2K_1\frac{R_c^5}{Q_c'} e^2 + \frac{8}{25}\left(1+\frac{57}{4}e^2\right)K_2\frac{R_\star^5}{Q_\star'}    \right] \nonumber\\
\label{Goldreich} \frac{1}{e}\frac{de}{dt} & = & -\frac{1}{a^{13/2}}\left[ K_1\frac{R_c^5}{Q_c'} + K_2\frac{R_\star^5}{Q_\star'} \right] \, ,
\end{eqnarray}
where $R_c$ and $R_\star$ are the companion and primary radii,
$Q'_{c}$ and $Q'_{\star}$ are the tidal dissipation factors of the
companion and the primary and $K_{1,2}$ are constants given by
\begin{eqnarray}
K_1 & = & \frac{63}{4}\left(GM_\star\right)^{1/2}\frac{M_\star}{M_c} \nonumber\\
K_2 & = & \frac{225}{16}\left(GM_\star\right)^{1/2}\frac{M_c}{M_\star} \, ,
\end{eqnarray}
where $M_c$ is the companion mass and $M_\star$ is the primary mass.

Furthermore, the radius and mass of the companion are held constant
while those of the primary are time-dependent.  Note that in these equations, $e$ can only decrease and if the primary mass were constant, $a$ could only decrease as well.

The tidal dissipation factors $Q'_c$ and $Q'_\star$ are dimensionless
parameters given by $Q'=3Q/2k_{2}$, where $Q$ is the specific tidal
dissipation function and $k_{2}$ is the Love number
\citep{Goldreich:1963nr, Goldreich:1966qv, Ogilvie:2007kb}.  $Q$ is a
dimensionless parameter that quantifies the efficiency of tidal
dissipation inside a body.  It is defined by its reciprocal, the
specific dissipation function
\begin{equation}
 Q^{-1}=\frac{1}{2 \pi E_{0}} \oint_{}^{} \left( -\frac{dE}{dt} \right) dt \, ,
 \label{eq:Q}
\end{equation}
where $E_{0}$ is the maximum energy stored in the tidal distortion of
the body and the integral $\oint_{}^{} \left( -\frac{dE}{dt}\right)
dt$ is the energy dissipated over one orbital cycle
\citep{Goldreich:1963nr}.  Typical estimates yield $Q'_\star \sim
10^{5.0-8.0}$, $Q'_c \sim 10^{5.5-6.5}$ for Jupiter mass planets
around main-sequence stars and $Q'_{\rm Jupiter}\sim10^{5.5-6.0}$
\citep{Goldreich:1966qv, Yoder:1981zv, Ogilvie:2004sy, Ogilvie:2007kb,
  Jackson:2009bq}

\subsection{Primary and Companion Tides}
\label{ssec:pctides}
Tides on the primary act to synchronize its spin while tides on the
companion act to circularize the orbit.  We can compare the tides
raised on the companion (induced by the primary) to the tides raised
on the primary (induced by the companion) as follows:
\begin{eqnarray}
\frac{\left( da / dt \right)_c}{\left( da / dt \right)_\star} & = & \left( \frac{7e^2}{1+\frac{57}{4}e^2} \right) \left( \frac{M_\star}{M_c}\right)^2 \left( \frac{Q'_\star}{Q'_c}\right) \left(  \frac{R_c}{R_\star} \right)^5 \\
\frac{\left( de / dt \right)_c}{\left( de / dt \right)_\star} & = & \frac{28}{25} \left( \frac{M_\star}{M_c}\right)^2 \left( \frac{Q'_\star}{Q'_c}\right) \left( \frac{R_c}{R_\star} \right)^5 \, .
\end{eqnarray}
For a $1$ $M_\odot$ AGB star (with radius $\sim$1 AU) and a 0.1 $M_\odot$ main sequence companion (with radius $\sim$0.1 $R_\odot$), with $e=0.1$, we have
\begin{eqnarray}
 \dot{a}_c / \dot{a}_\star  & \sim & 10^{-16}\left(Q'_\star / Q'_c \right)\\
 \dot{e}_c / \dot{e}_\star & \sim & 10^{-15} \left( Q'_\star / Q'_c \right) \, .
\end{eqnarray}
For lower mass companions, these ratios remain extreme.  Therefore,
for $Q'_\star \simeq Q'_c$, if the primary is a post-main sequence
star, the orbit decays almost entirely due to tides raised on the
star, not to those raised on the companion.

\begin{figure}
\includegraphics[width=3.20in]{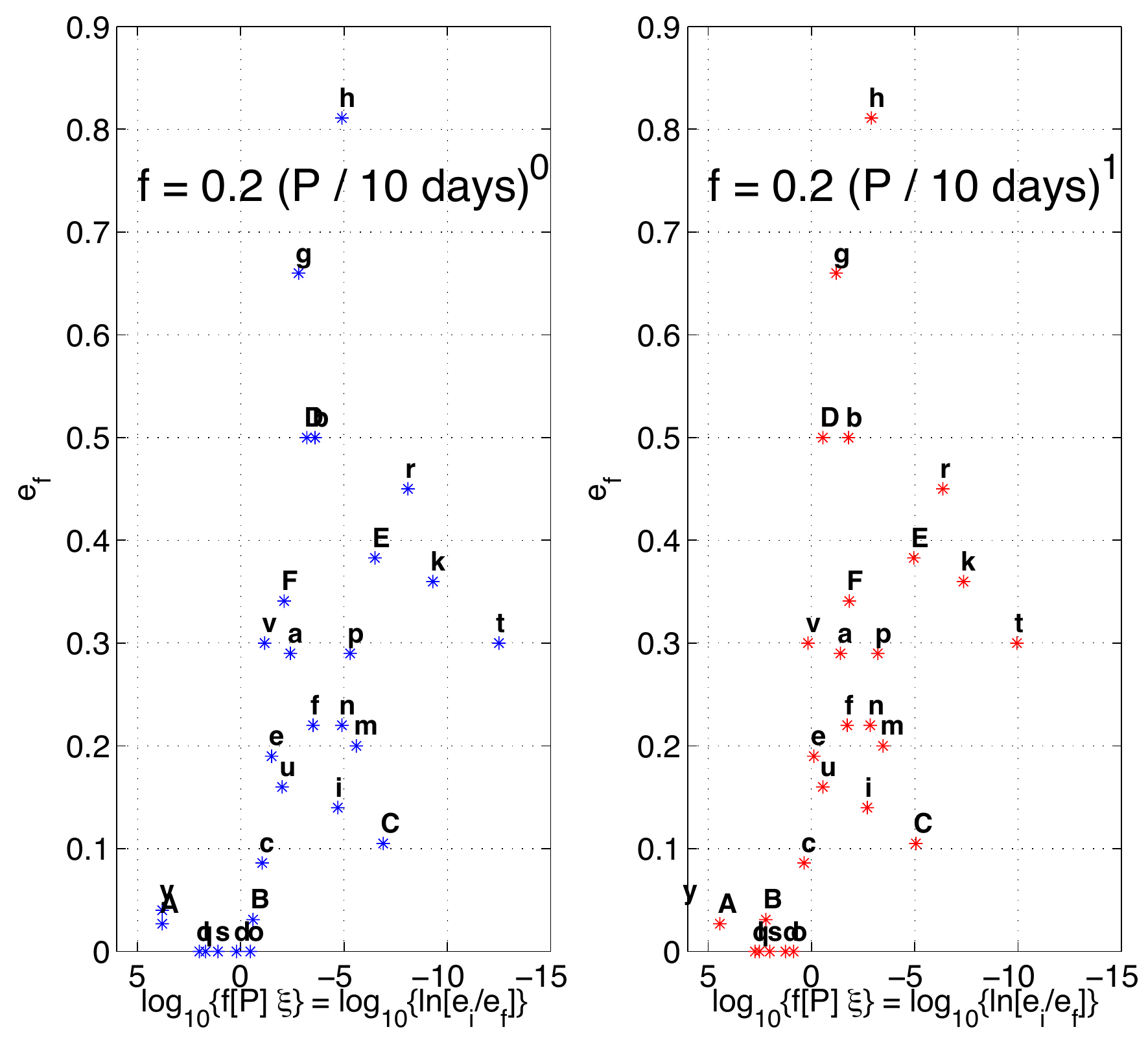}
\caption{Reproduced versions of Fig. 4c of \citep{Verbunt:1995rt}
  assuming a period dependence on $f$.  The transition between
  circularized and non-circularized systems is preserved around unity
  for a variety of scalings.  Here we present one example.  Left:
  $f\propto P^{0}$ which implies $Q'_\star\propto P$.  Right:
  $f\propto P^{1}$ which implies $Q'_\star\propto P^0$.
\label{VP_binaries}}
\end{figure}

As the orbit circularizes ($e \rightarrow 0$), the tides on the
companion decrease until they vanish for perfectly circular orbits.
However, note that some dissipation must occur within the companion to
maintain synchronous rotation.  In this limit, the orbital evolution
is governed by non-synchronous rotation between the companions orbit
and spin of the primary.  As a result, in-spiral continues and might
lead to the companion plunging into the primary (the cases we are
interested in here).

\subsection{Main-sequence and Post-MS Tides}
\label{ssec:MSPMStides}
For the known extrasolar giant planets (EGPs), the tides raised on the
primary can be substantial and can lead to significant orbital
reduction during the main-sequence \citep{Ibgui:2009jk,
  Levrard:2009qy, Ibgui:2009fk,Spiegel:aa}.  For the same $Q'_\star$,
we can estimate whether a similar reduction is expected during the
post-MS.  From the previous section, we have that the tides on the
primary dominate and, hence,
\begin{eqnarray}
\frac{\Delta {\rm ln} a_{\rm rgb}}{\Delta {\rm ln} a_{\rm ms}} & \sim & \left( \frac{R_{\rm rgb}}{R_{\rm ms}}\right)^5 \left( \frac{a_{\rm rgb}}{a_{\rm ms}}\right)^{-13/2}\label{eq:tidestrength} \\
 & & \times \left( \frac{M_{\rm c,rgb}}{M_{\rm c,ms}}\right) \left( \frac{\tau_{\rm rgb}}{\tau_{\rm ms}}\right)\left(\frac{Q'_{\star,rgb}}{Q'_{\star,agb}}\right)^{-1}\nonumber \, ,
\end{eqnarray}
where $\tau_{\rm ms}$ is the typical lifetime during the main sequence
and $\tau_{\rm rgb}$ is the lifetime during the RGB phase.

For the main sequence, we take a 1 Jupiter mass companion around a
1~$M_\odot$ main-sequence primary with semimajor axis $\sim$0.05~AU.
For the RGB phase, we assume that the companion is orbiting at 1.3~AU,
that $Q'_{\rm \star,ms} = Q'_{\rm \star,rgb}$ and that $R_{\rm
  rgb}\sim 10^2 R_{\rm ms}$ and $\tau_{\rm rgb}\sim 10^{-2} \tau_{\rm
  ms}$.  Using these parameters yields $\Delta {\rm ln} a_{\rm rgb} /
\Delta {\rm ln} a_{\rm ms} \sim0.063$.  Therefore, for this system, we
expect tides to be weaker on the RGB phase than they are for EGP
systems.  This is likely to be true for low-mass primaries ($\lesssim
1.5 M_\odot$) where radial expansion during the RGB is substantial
(see Fig.~\ref{models}).  For higher mass primaries, radial expansion
is minimal during the RGB and extensive during the AGB.  However,
typical AGB lifetimes are $\sim$$10^{-2} - 10^{-3} \tau_{\rm rgb}$.
We can estimate the effect of tides during the AGB phase in the same
way.  Assuming $R_{\rm agb} \sim200$ $R_\odot$, $Q'_{\rm \star,ms} = Q'_{\rm
  \star,agb}$ and $\tau_{\rm agb} \sim10^{-5} \tau_{\rm ms}$, we have
that $\Delta{\rm ln} a_{\rm agb} / \Delta {\rm ln} a_{\rm ms}
\sim2.0\times10^{-3}$.  Thus in general, we expect synchronization
tides on the AGB to be weaker both than those on the RGB and for
typical EGPs around main sequence stars.  Note the steep dependence on the ratio of the stellar radii in Eq. \ref{eq:tidestrength}.  If $R_{\rm agb}=2R_{\rm rgb}$ and $\tau_{\rm agb}=(1/30)\tau_{\rm rgb}$, then $\Delta{\rm ln} a_{\rm agb} / \Delta {\rm ln} a_{\rm rgb}\sim1$ and tides are approximately equal during the RGB and AGB phases.  However, as will be seen shortly, there is reason to expect that $Q'_*$ may be orders of magnitude smaller on the giant branches than on the main sequence for the same $a/R_*$, with a corresponding increase in the importance of the advanced evolutionary phases for the orbit.

\subsection{A Period Dependence of $Q'_\star$?}
\label{ssec:Q*[P]}
It is worth nothing that the work of \citet{Goldreich:1963nr} and
\citet{Goldreich:1966qv} does not specify a tidal dissipation
mechanism.  \citet{Zahn:1966jk} however, proposed a tidal theory based
on turbulent viscosity.  This theory was tested and calibrated in
stellar binaries that contain an evolved star primary
\citep{Verbunt:1995rt}.  Evolved stars are expected to possess
extended convective zones, which could be crucial for tidal
dissipation.  In their formalism, the authors introduce a
dimensionless factor $f$ that is calibrated via the orbital
eccentricity measurements in their post-MS binary sample.  The authors
argue that observational data imply that $f$ is constant and
approximately unity.

By setting equal the expression for the tidal torque (as a function of $Q'_\star$) in \citet{Goldreich:1966qv} to the expression (as a function of $f$) in \citet{Zahn:1989lr}, as simplified by \citet{Verbunt:1995rt}, we have that

\begin{eqnarray}
Q'_\star & = & \frac{63}{16\pi}\frac{GM_\star}{R_\star^3}\frac{M_\star}{fM_{\rm env}}\left( \frac{M_{\rm env}R_\star^2}{L} \right)^{1/3}P\label{Zahn}\\
 & \propto &\frac{\tau_{\rm conv}}{\tau^2_{\rm dyn}}P\nonumber
\end{eqnarray}
where $M_{\rm env}$ is the mass of the convective envelope, $P$ is the
orbital period, $\tau_{\rm conv}$ is the convective timescale and
$\tau_{\rm dyn}$ is the dynamical timescale.  If
$f$ is in fact constant, then $Q'_\star$ is proportional to orbital
period.  Note that the numerical prefactor in Eq. (10) ($63/16\pi$) is slightly different from the $225/32\pi$ that results from setting our expresion for $d\ln e/dt$ (in Eq.~2) equal to the corresponding expresion in  \citet{Verbunt:1995rt}.  The difference (compared to the variation in period) is sufficiently small that it does not effect our results.  

The argument for $f$ constant and $\sim$1 derives from Fig.~4c of
\cite{Verbunt:1995rt}.  The transition between circularized and
non-circularized systems occurs when the abscissa is 0.  The abscissa
is given by ${\rm log_{10} \left\{-\Delta {\rm ln \left[e\right]} / f
  \right\}} = {\rm log_{10}} \left\{ {\rm ln \left[e_i / e_f \right]
}\right\} - {\rm log_{10} \left\{f\right\} }$.  If ${\rm log_{10}}
\left\{ {\rm ln \left[e_i / e_f \right] }\right\}$ is greater than
(less than) 0, the system is strongly (barely) circularized.  The
transition occurs near ${\rm log_{10}} \left\{ {\rm ln \left[e_i / e_f
    \right] }\right\} = 0$.  Looking at Fig.~4c of
\cite{Verbunt:1995rt}, we see this transition happens at ${\rm
  log_{10}} \left\{ {\rm ln \left[e_i / e_f \right] }\right\} - {\rm
  log_{10} \left\{f\right\} } \sim0$, therefore, $f\sim1$.

While this argument does constrain $f$ for systems at the sharp
transition between circularization and non-circularization, it does
not rule out a dependence of $f$ on period.  In fact, the transition
between circularized and non-circularized systems remains at
$\log\left\{\Delta\ln[e]\right\} \sim 0$ if $f$ scales as $P^{1-x}$
with $1\gtrsim x\gtrsim 0$ and has an appropriate normalization (see
Fig.~\ref{VP_binaries}).  This corresponds to $Q'_\star \propto P^x$.
It is worth mentioning that the scaling used in Fig.~\ref{VP_binaries}
is one example among many that preserve the location of the transition.
Different normalizations preserve the transition for different ranges
of $x$.  For example, $f =\left(P / 200\, {\rm days} \right)^{1-x}$,
preserves the transition for $2\gtrsim x \gtrsim 0$.  The value $x=2$ (for which $Q'_\star\propto P^{-1}$) has been urged by \cite{Goldreich:1977qy} in the limit $P\ll \pi\tau_{\rm conv}$; however, $P\gtrsim\pi\tau_{\rm conv}$ for most of the cases of interest to us.


Thus, by using Eq.~(\ref{Zahn}), we are able to employ the tidal
prescription of \citet{Zahn:1966jk} with the observational calibration
of \cite{Verbunt:1995rt}.  In Section 4, we show that employing such a
formalism leads to values of $Q'_\star$ between $\sim$$10^2$ and
$10^3$ for post-main sequence giants.  These values are two to seven
orders of magnitude lower than those typically invoked in the context
of extra-solar giant planets \citep{Goldreich:1966qv, Yoder:1981zv,
  Ogilvie:2004sy, Ogilvie:2007kb, Jackson:2009bq, Ibgui:2009jk,
  Ibgui:2010lr, Miller:2009lr} and lead to strong tidal interactions.
We pause to note that tidal theory is an active field of research
\citep{Goodman:2009kx, Gu:2009yq, Arras:2009rt} Particularly relevant is
a recent study that investigated a frequency dependence of $Q'$
\citep{Greenberg:2009lr}.  This work demonstrated that commonly used
analytic approaches (which make analogy to a driven harmonic
oscillator) are probably valid only for low eccentricities and
inclinations.  Theoretical and observational constraints on the tidal
dissipation mechanism will help constrain the results of this work in
the future.

For now, in light of the uncertainties in $Q'_\star$, and in $f$, we
adopt the following form
\begin{equation}
Q'_\star =  Q_0\times\left( \frac{\Pi}{\Pi_0}\right)^x
\label{eq:Jeremy}
\end{equation}
where $\Pi\equiv P/\tau_{\rm dyn} = P\left(\rho G\right)^{1/2}$ is the
orbital period divided by the dynamical time, $\Pi_0$ is a reference
value, $\log_{10} Q_0$ is between 5 and 9.  For $\Pi_0$, we use a
10~day orbital period divided by the dynamical time of the Sun, namely
$\Pi_0 = (10~{\rm days})\times \left(\rho_\odot G \right)^{1/2}$.
Note that even in the case where $f=1$ and $x=1$,
Eq.~(\ref{eq:Jeremy}) is not fully consistent with Eq.~(\ref{Zahn}).
For this paper, we focus on two representative values of $x$ such that
$Q'_\star\propto \Pi^0$ and $Q'_\star\propto \Pi^1$.

\section{Primary and Companion Models}
\label{sec:PCmodels}
\begin{figure}
\begin{center}
\includegraphics[width=8.5cm,angle=0,clip=true]{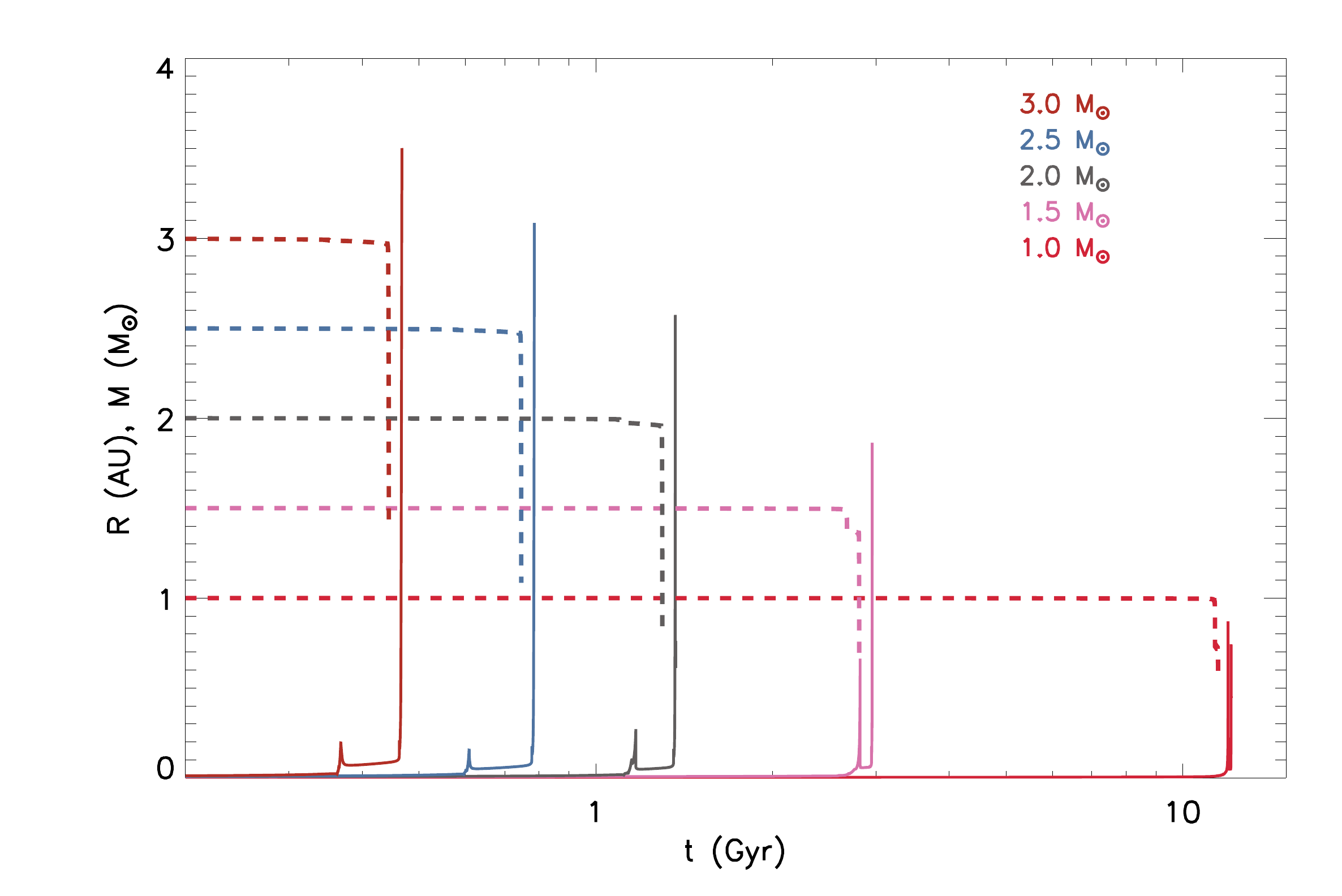}
\end{center}
\caption{Mass (dotted line) and radius (solid line) profiles for a
  subset of our stellar models (i.e. those with $\eta=1$).  The mass
  profiles have been offset 5\% in time so that the evolution is
  distinguishable on the plot.
\label{models}}
\end{figure}

Our stellar evolution models are calculated using the ``Evolve
Zero-age Main Sequence (EZ) code" \citep{Paxton:2004wd}.  For each
initial stellar mass, we evolve the star from the ZAMS through the end
of the AGB phase including a Reimers mass-loss prescription
\citep{Reimers:1975lr}.  In order to explore the influence of the rate
of mass-loss on our calculations, we use a range of values of the
Reimers $\eta$ parameter: 0.7, 1, and 5.  Each stellar model has a
metallicity of $Z=0.02$.  In addition to the stellar radius and mass,
we calculate the core mass and envelope binding energy as a function
of time.  Our stellar models (for $\eta = 1$) are presented in
Fig.~\ref{models} while key parameters for all models are summarized
in Table~\ref{table1}.

\subsection{Companions}
\label{ssec:Companions}
We consider three types of companions: planets ($M_c \lesssim 0.0026
M_\odot$; \citealt{Zapolsky:1969hs}), brown dwarfs ($0.0026
M_\odot\lesssim M_c \lesssim 0.077 M_\odot$; \citealt{Burrows:1993zv})
and low-mass main sequence stars ($0.077 \lesssim M_c \lesssim 0.1
M_\odot$).  We adopt the following approximation to the models of
\citet{Burrows:1993zv, Burrows:1997lr, Burrows:2001fk} for brown dwarf
radii
\begin{eqnarray}
R_2 & = & \left(0.117-0.054{\rm log_{10}}^2\left[\frac{M_c}{0.0026}\right] \right. \\
\nonumber & & \left. +0.024{\rm log_{10}}^3\left[\frac{M_c}{0.0026}\right]\right)R_\odot \, ,
\end{eqnarray}
and use a homologous power-law for low-mass stellar radii
\begin{equation}
R_2=\left(\frac{M_c}{M_\odot}\right)^{0.92}R_\odot \, ,
\end{equation}
\citep{Reyes-Ruiz:1999lr}.

We carry out our calculations for the following companions: a
0.001~$M_\odot$ (1~$M_{\rm J}$) planet, a 0.01~$M_\odot$ (10~$M_{\rm
  J}$) object that could be either a massive planet or a low-mass brown-dwarf (depending on formation scenario; hereafter referred to as a brown dwarf), and a 0.1~$M_\odot$ (100~$M_{\rm J}$) low-mass main
sequence star.

\begin{table*}
 \centering
 \begin{minipage}{140mm}
  \begin{tabular}{@{}llllllllllllr@{}}
  \hline
    &    $M_{\rm zams}$ & $R_{\rm rgb}$\footnote[1] {Maximum radius on the RGB/AGB.} & $M_{\rm rgb}$& $R_{\rm agb}$ & $M_{\rm agb}$ &  $M_{\rm core,rgb}$\footnote[2]{Calculated when the stellar radius is largest in the RGB and AGB phases.} \footnote[3]{The core masses are defined by composition rather than degeneracy.  Explicitly, it is the mass interior to the outermost hydrogen-free shell.} & $M_{\rm core,agb}$\\
&   ($M_\odot$) & ($R_\odot$) & (AU) & (AU) 
     &  & ($M_\odot$) & ($M_\odot$) \\
 \hline
 \hline
  &1.00&0.82 &0.84&1.06 &0.60& 0.45 & 0.58 \\
 $\eta = 0.7$& 2.00 & 0.27 &1.98 & 2.81 &0.90& 0.38 &  0.81 \\
  &3.00 & 0.20 &2.99 & 3.71 &1.60& 0.43 & 1.07  \\
  \hline
  \hline
  &1.00&0.87 &0.73 &0.75 &0.56& 0.46 & 0.56 \\
 & 1.50 &0.67 &1.38 & 1.85 &0.69& 0.45 &  0.69 \\
 $\eta = 1$& 2.00 & 0.27 &1.97 & 2.57 &0.80& 0.39 &  0.81 \\
  &2.50 & 0.16 & 2.49 & 3.14 &1.07& 0.38 & 0.91  \\
  &3.00 & 0.20 &2.98 & 3.50 &1.40& 0.44 & 0.99  \\
  \hline
   \hline
  &1.00\footnote[4]{The $\eta=5$, $1$ $M_\odot$ model does not achieve a sufficient core mass to undergo a helium flash.}&0.41 &0.38 & -- & -- &  0.37 & -- \\
 $\eta = 5$& 2.00 & 0.30 &1.84 & 1.37 &0.66& 0.39 &  0.60 \\
  &3.00 & 0.20 &2.94 & 2.50 &0.81& 0.43 & 0.75 \\
  \hline

\end{tabular}
\end{minipage}
 \caption{Stellar model data.  Column quantities from left to right:
   zero-age main sequence mass, maximum
   radius on the RGB, stellar mass at the tip of the RGB, maximum radius on the AGB, stellar mass at the tip of the AGB, core mass for the corresponding
   $R_{\rm rgb}$, core mass for the corresponding $R_{\rm agb}$.
\label{table1}}
\end{table*}

\section{Tidal Results}
\label{sec:TResults}
As the binary system evolves, mass-loss and tidal torques are in
competition.  Mass lost from the system acts to increase the semimajor
axis while tidal torques decrease it.  For each primary and companion,
we compute the evolution of the orbit from the zero-age main sequence
through the post-main sequence.  If the companion is tidally captured
(i.e. plunges into the primary star), it enters a common envelope with
the primary and we halt the calculation of the orbital evolution.  If
the companion evades tidal engulfment, mass-loss continues and the
orbit expands until the end of the evolutionary model.

We performed calculations for two tidal prescriptions: those
corresponding to Eq.~(\ref{Goldreich}) and those corresponding to
Eq.~(\ref{Zahn}) when $f=1$.  With this approach, we can directly
compare the tidal theories of \citet{Goldreich:1966qv} to those of
\citet{Zahn:1966jk} when $f$ is constant and equal to one
\citep{Verbunt:1995rt}.

For each stellar model, companion mass, and tidal theory, we calculate
the maximum initial semimajor axis, $a_{\rm i,max}$, that is tidally
engulfed.  Eq.~(\ref{Goldreich}) is accurate to lowest order in
eccentricity and valid for $e \ll 1$ \citep{Goldreich:1966qv}.  For
low $e$ ($\lesssim$0.1), the effect of eccentricity is to modify the dominant tidal
term by $\lesssim$10\%.  Thus, we assume circular orbits for the
ensuing calculations.  It should be noted that the formalism we employ only 
considers the quadrupole component of the tidal potential.

Although the Reimers prescription is an imperfect description of
mass-loss, we investigate how different rates of mass-loss influence
our results by using $\eta$ values of 0.7, 1, and 5
\citep{Reimers:1975lr}.  Our mass-loss prescription reproduces typical
mass-loss rates during the RGB but underestimates mass-loss rates on the
AGB.  As such, the results presented here serve as an approximate
upper limit to the parameter space where tidal torques dominate over
mass loss.  Access to accurate stellar evolution models and realistic
mass-loss prescriptions (motivated by observations during all phases
of stellar evolution) will refine this work in the future
\citep{Schroder:2005uq, Schroder:2007qy}.

\subsection{When $f = 1$}
\label{ssec:f=1}
The tidal theory of \citet{Zahn:1966jk} under the assumption that
$f=1$ has been employed to determine the effect of tidal torques on
various aspects of post-MS evolution \citep{Soker:1995lr,
  Soker:1996fk, Villaver:2009qy, Carlberg:2009uq, Bear:2009yq}.  This
is accompanied by results that suggest strong tidal interactions for
companions within $\sim$$(5-7)\times R_{\rm max}$, where $R_{\rm max}$
is the maximum radial extent of the star during either the RGB or AGB
phases \citep{Soker:1995lr, Soker:1996fk, Debes:2002kx,Moe:2006fc}.  Here, we make
the same assumption and determine the maximum orbital separation that
plunges into the primary star for various primary and companion
masses.

\begin{figure}
\begin{center}
\includegraphics[width=8.5cm,angle=0,clip=true]{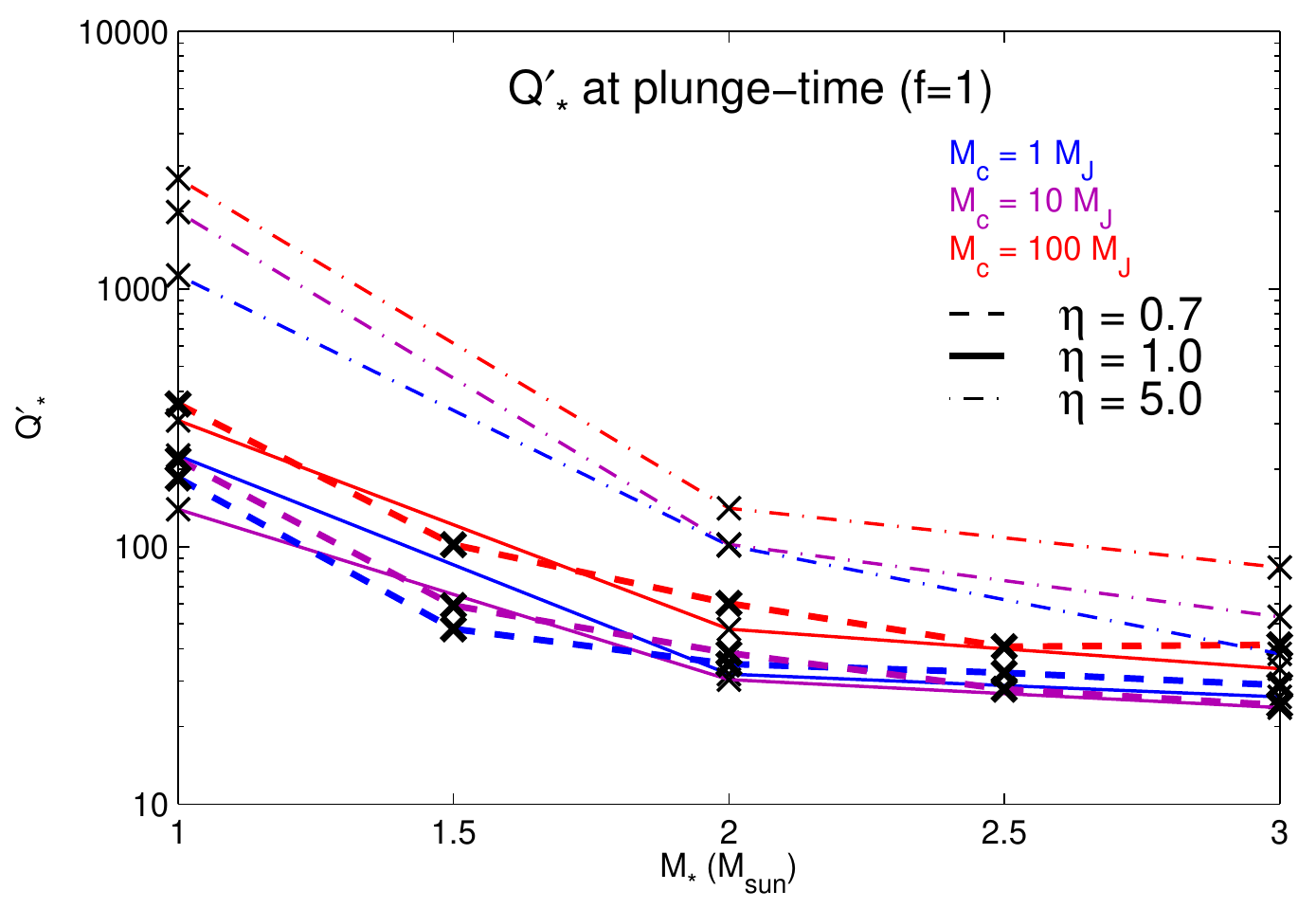}
\end{center}
\caption{$Q'_\star$ values for post-main sequence binaries if $f = 1$.
  Note that choice of $f = 1$ yields $Q'_\star$ values 2 to 7 orders
  of magnitude lower than typical values used to explain the orbits
  and radii of transiting extra solar planets.
\label{Q_f_1}}
\end{figure}

Figure~\ref{Q_f_1} shows $Q'_\star$ values at the time of tidal
engulfment under the assumption that $f=1$.  Typically, values range
from $10^{1.3} \lesssim Q'_\star \lesssim 10^{3.5}$, and, for a given
progenitor mass, $Q'_\star$ values vary by less than a factor of
$\sim$20, irrespective of the choice of $\eta$ and companion mass.
This is striking, as it implies that the assumption of $f=1$ leads to
tidal dissipation rates that are $\sim$2-7 orders of magnitude larger, in a dimensionless sense, than those acting
on extra-solar giant planets.  Typical $Q'_\star$ values for the
transiting EGPs are typically taken to range from $10^5 - 10^8$
\citep{Barnes:2008lr, Jackson:2009bq, Ibgui:2009jk, Ibgui:2010lr,
  Miller:2009lr, Levrard:2009qy}.

\begin{figure}
\begin{center}
\includegraphics[width=8.5cm,angle=0,clip=true]{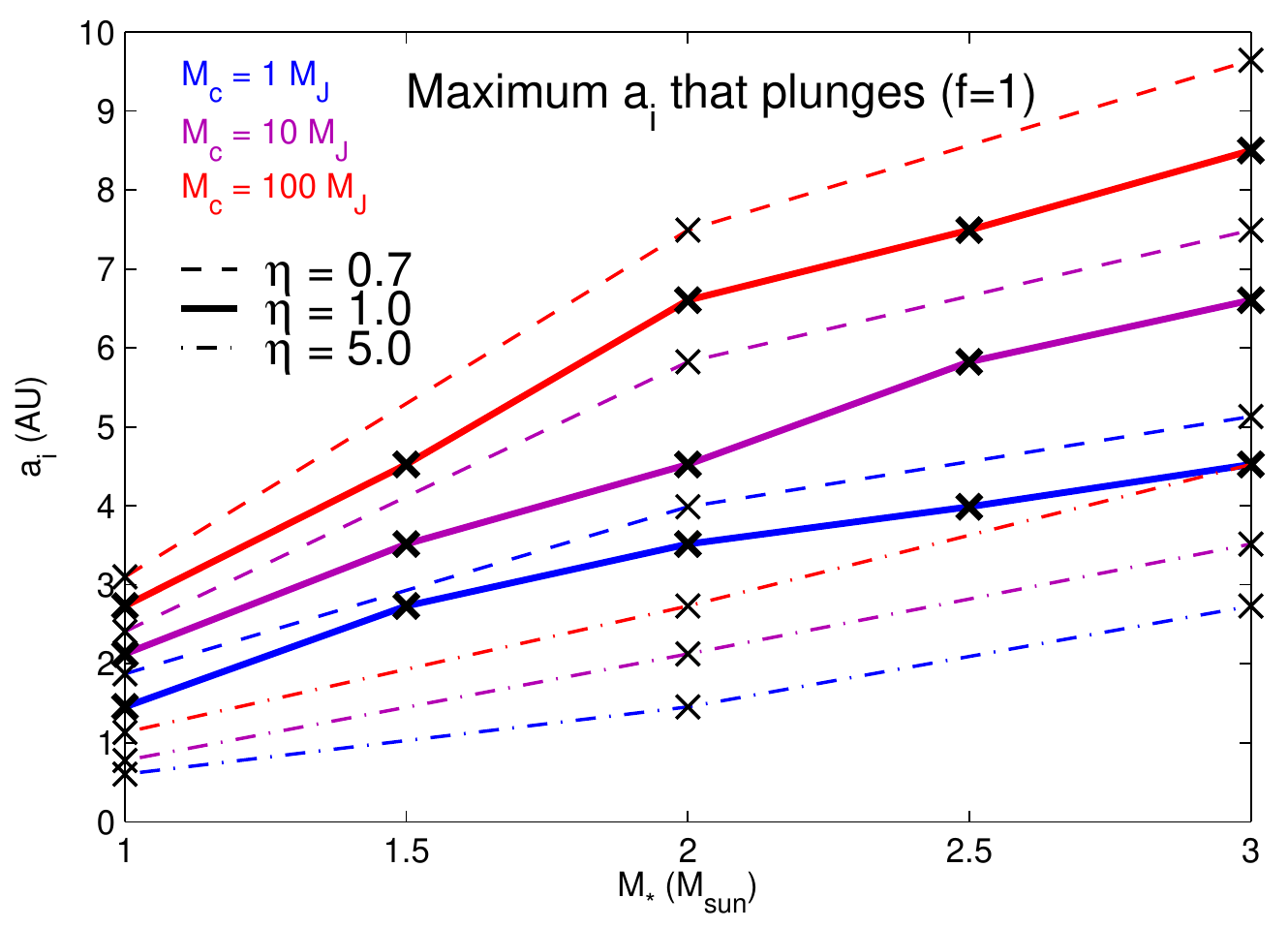}
\vspace{0.1cm}
\includegraphics[width=8.5cm,angle=0,clip=true]{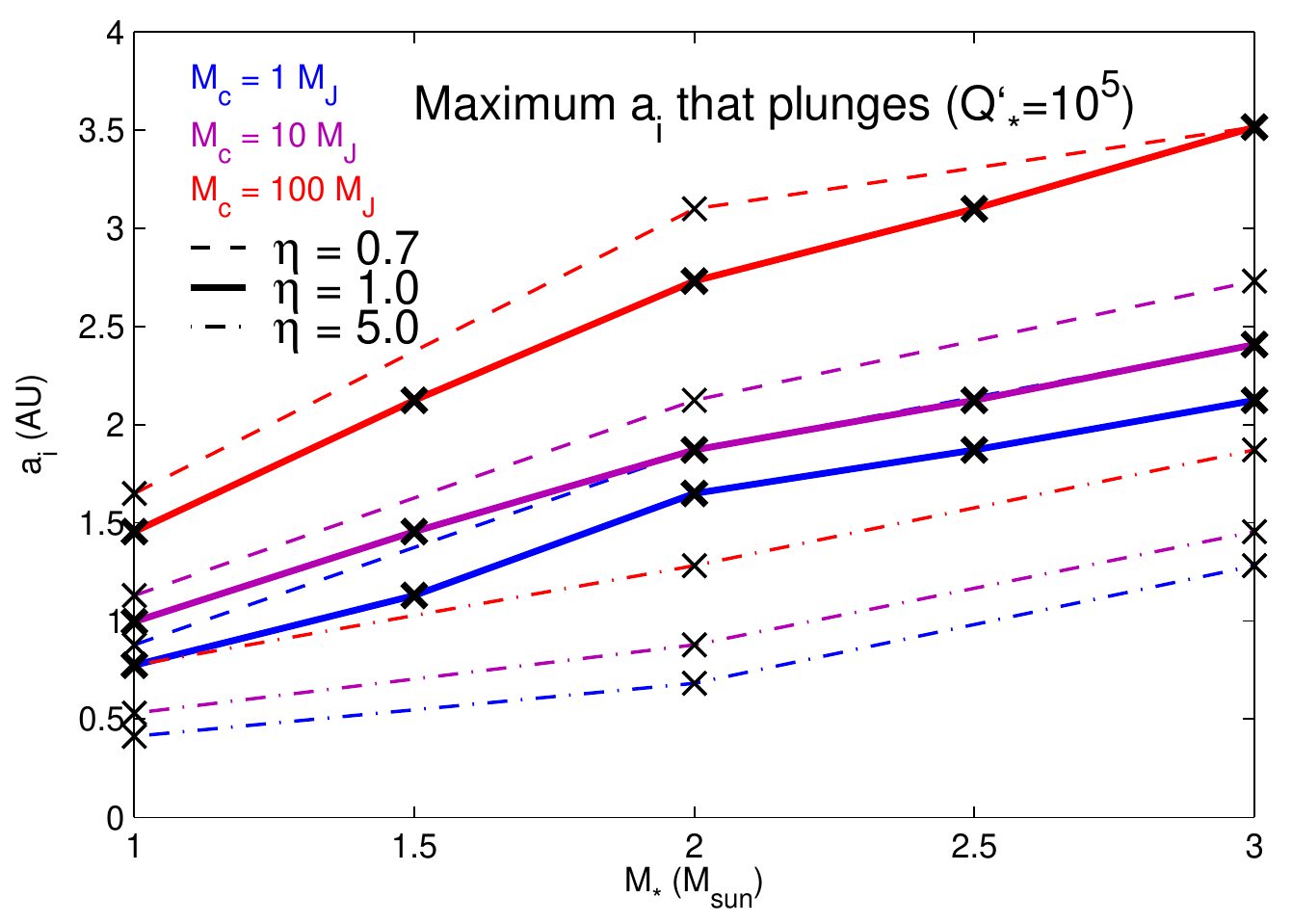}
\end{center}
\caption{Maximum semimajor axis, $a_{\rm i,max}$ that is tidally
  engulfed under the assumption that $f=1$ (top) and $Q'_\star = 10^5$
  (bottom).  Both figures present results for companions of mass $1
  M_{\rm J}$ (blue curves), $10 M_{\rm J}$ (purple curves) and $100
  M_{\rm J}$ (red curves).  Results for different mass-loss models are
  shown by $\eta=0.7$ (dashed curves), $\eta=1$ (solid curves),
  $\eta=5$ (dot-dashed curves).
\label{f_1}}
\end{figure}

With small $Q'_\star$ values, one would expect tidal engulfment to
occur at large separations.  Figure~\ref{f_1} shows the maximum
semimajor axis that is tidally captured ($a_{\rm i,max}$) for a range
of primary models and companion masses.  The top panel presents
$a_{\rm i,max}$ in units of AU for the case of $f=1$, while the bottom panel shows the same results for the case of $Q'_\star=10^5$ (see Table 1).  Generally,
the more massive stellar models extend to larger radii on the AGB and,
therefore, can swallow companions that are farther away.  Jupiter-mass
companions within $\lesssim$2$\times R_{\rm max}$ are captured while
more massive companions can plunge at farther distances
(10-Jupiter-mass companions are captured within $a \lesssim 2.5 \times
R_{\rm max}$; 100-Jupiter-mass companions are captured within
$a \lesssim 3 \times R_{\rm max}$).  It should be noted that the time of tidal engulfment roughly corresponds to the time at which the primary star is at its maximal radial extent.  However, this is not exact as most binary systems are engulfed when the primary is still expanding, but near maximum radial extent.

If tides acting during the post-main sequence are characterized by
similar $Q'$ values (in the \citealt{Goldreich:1966qv} formalism) to
those acting on EGPs during the main sequence (i.e.,
$Q'_\star\gtrsim10^5$, $Q'_c\sim10^6$), then results might be
significantly different from those derived from the formalism of
\cite{Zahn:1966jk} with $f=1$ (in Fig.~\ref{f_1}).  To compare, in the
next section we present cases in which $Q'_\star$ is constant and
$Q'_\star\propto\Pi$.

\subsection{$Q'_\star$ constant and $Q'_\star\propto\Pi$}
\label{ssec:Qdepend}
\begin{figure}
\begin{center}
\includegraphics[width=4.0cm,angle=0,clip=true]{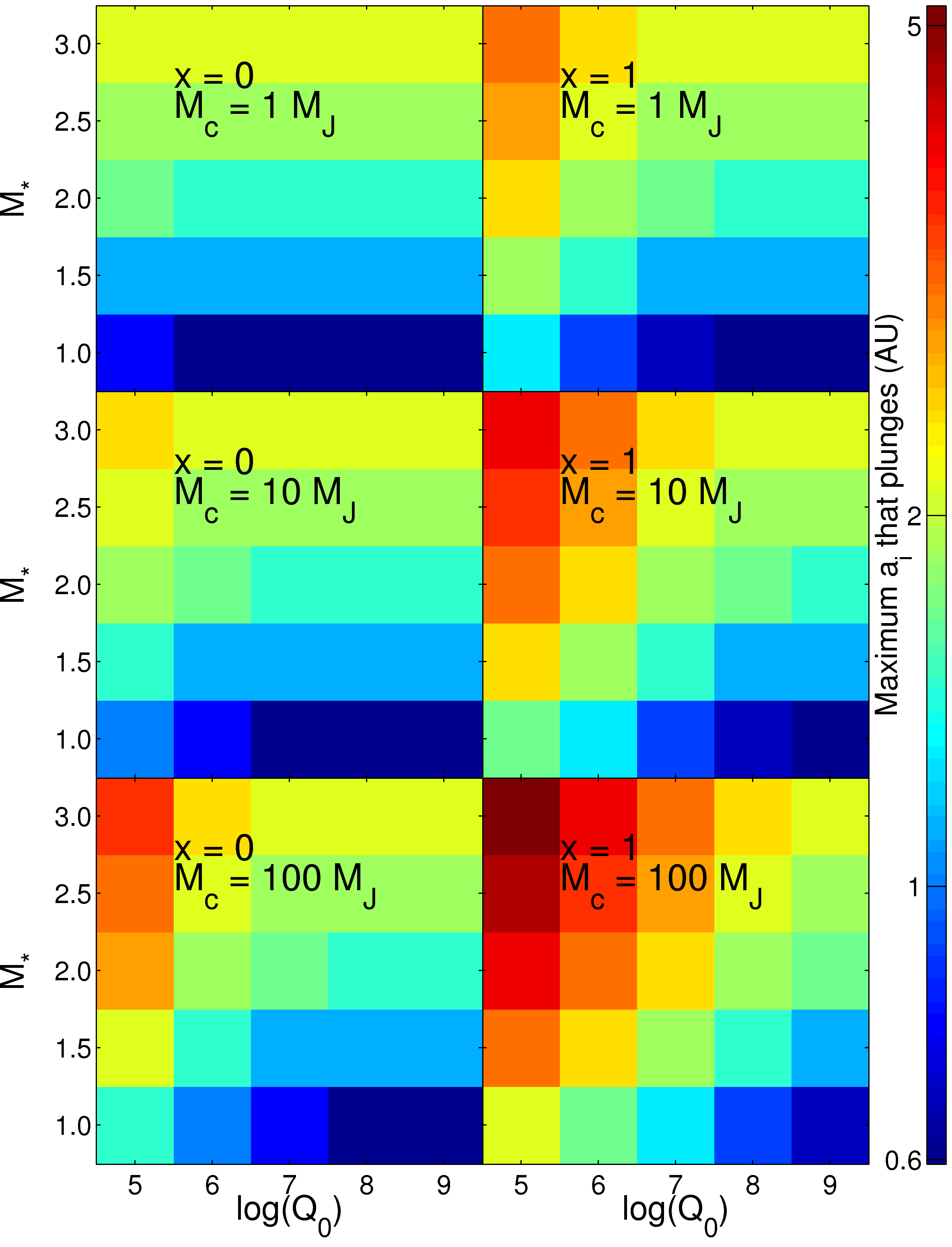}
\includegraphics[width=4.0cm, angle=0,clip=true]{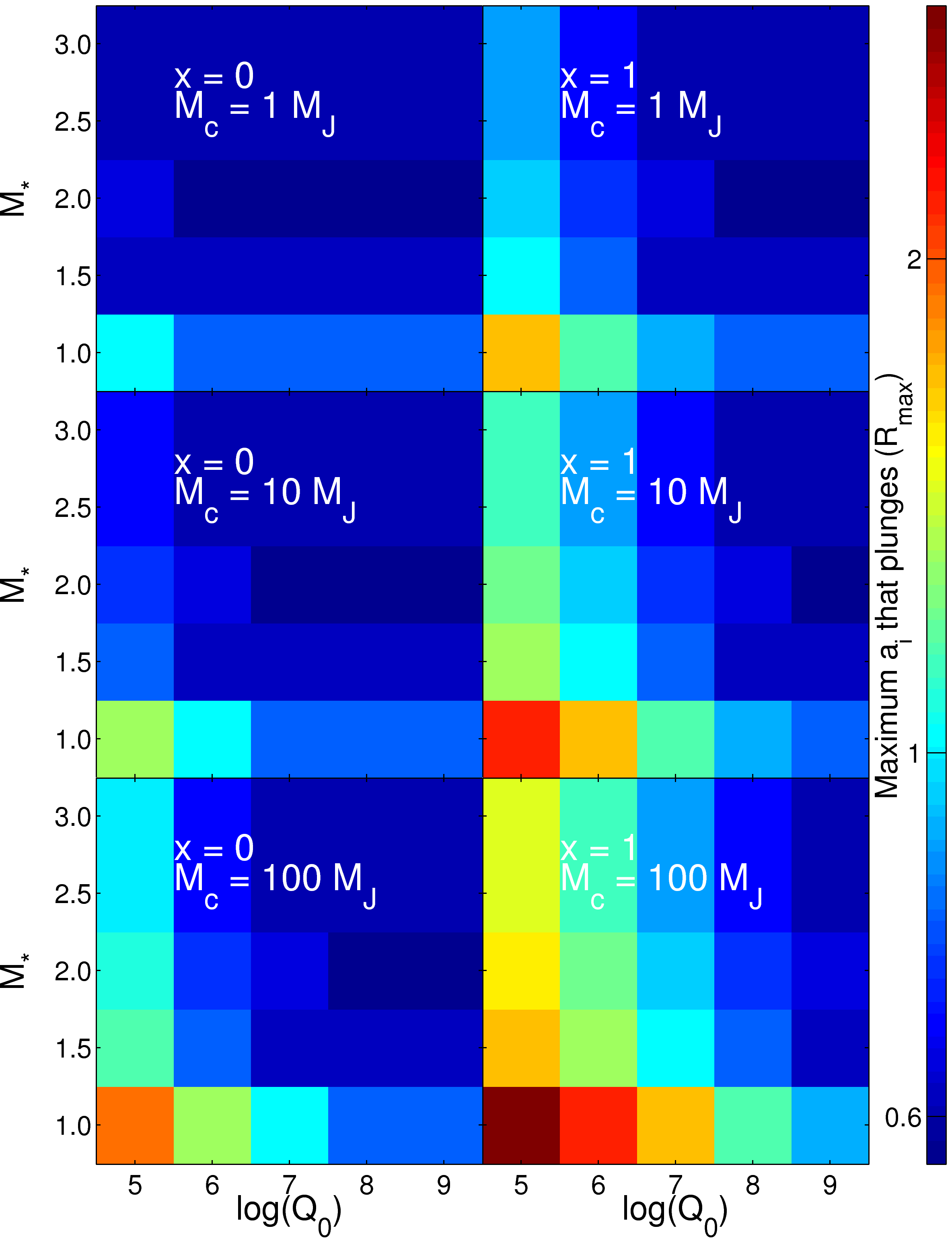}
\end{center}
\caption{Maximum initial semimajor axis, $a_{\rm i,max}$, that enters
  a common envelope for 1, 10 and 100~$M_{\rm J}$ companions with 1,
  1.5, 2, 2.5 and 3~$M_\odot$ primary stars.  The left column in each
  figure presents results for $x = 0$ [$Q'_\star = Q_0$] while the
  right column presents results for $x = 1$ [$Q'_\star = Q_0\times(\Pi
    / \Pi_0)$].  The evolutionary models of the primary star were
  computed with a Reimers mass-loss parameter of $\eta = 1$. The left
  figure is in units of AU while the right figure is in units of the
  maximum radius on the AGB (see Table 1).
\label{amax_eta1}}
\end{figure}

The left columns of Fig.~\ref{amax_eta1} show $a_{\rm i,max}$ as a
function of primary mass for constant $Q'_\star$.  The right columns
present $a_{\rm i,max}$ for $Q'_\star = Q_0 \times (\Pi / \Pi_0)$.
The evolutionary models in Fig.~\ref{amax_eta1} were calculated with a
Reimers mass-loss parameter of $\eta=1$.  The left plot is in units of
AU while the right plot is in units of $R_{\rm max}$.  As such, we can
directly compare the results of a constant $Q'_\star$ to those of
$f=1$ shown in Fig.~\ref{f_1}.  Note that for $10^5\leq Q'_\star \leq
10^9$ tidal torques are much weaker and result in smaller $a_{\rm
  i,max}$.  In addition, the ratio $a_{\rm i,max}/ R_{\rm max}
\lesssim 1$ for most of the parameter space.  For the same companion
and stellar evolution model, $Q'_\star \propto \Pi$ produces larger
$a_{\rm i,max}$ than the constant $Q'_\star$ case.  Finally, for the
EGPs, it may be that $x<0$, consistent with the prediction that $x$
should switch from $+1$ to $-1$ at $P\ll\tau_{\rm conv}$
\citep{Goldreich:1977qy}.

\begin{figure}
\begin{center}
\includegraphics[width=4.0cm,angle=0,clip=true]{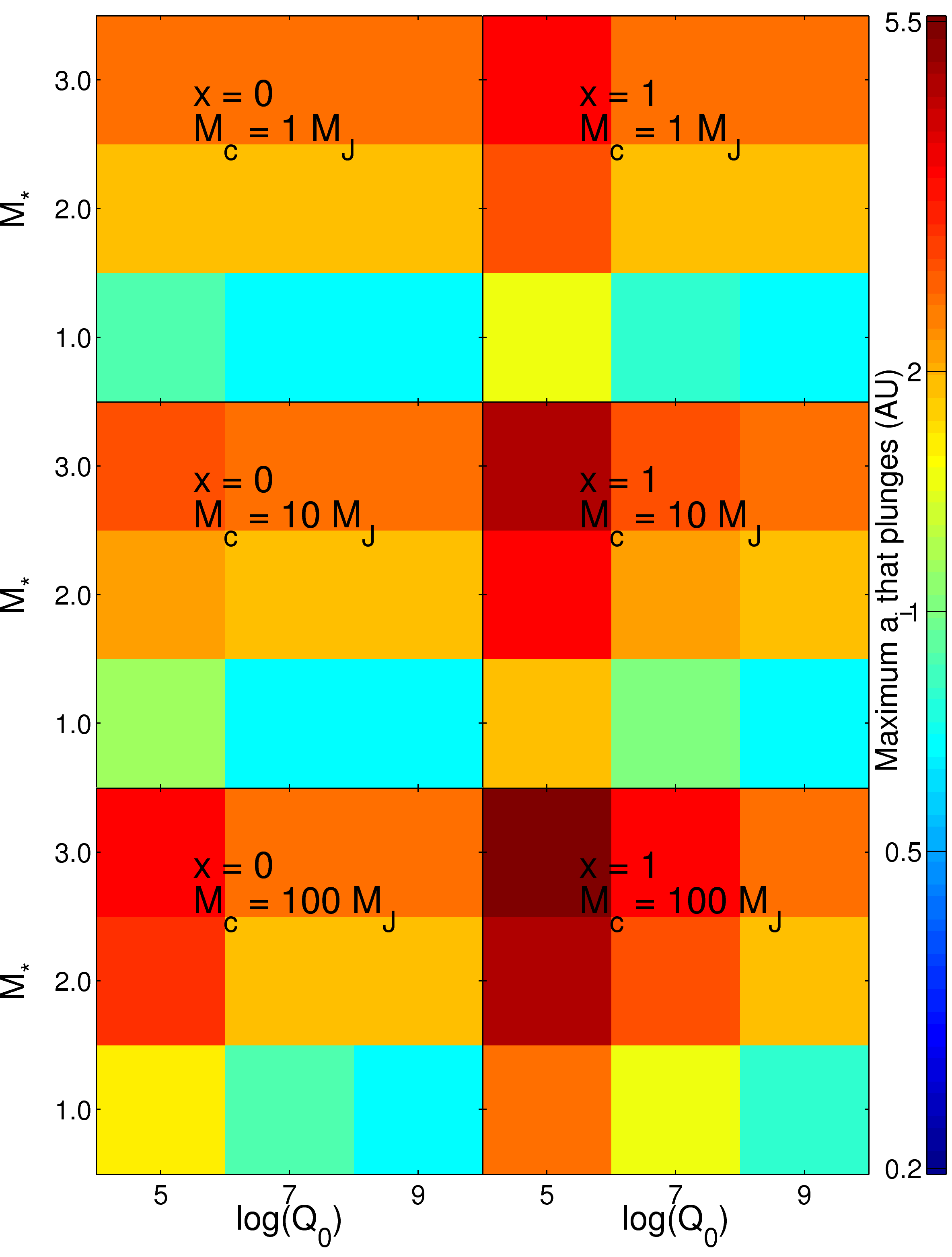}
\includegraphics[width=4.0cm, angle=0,clip=true]{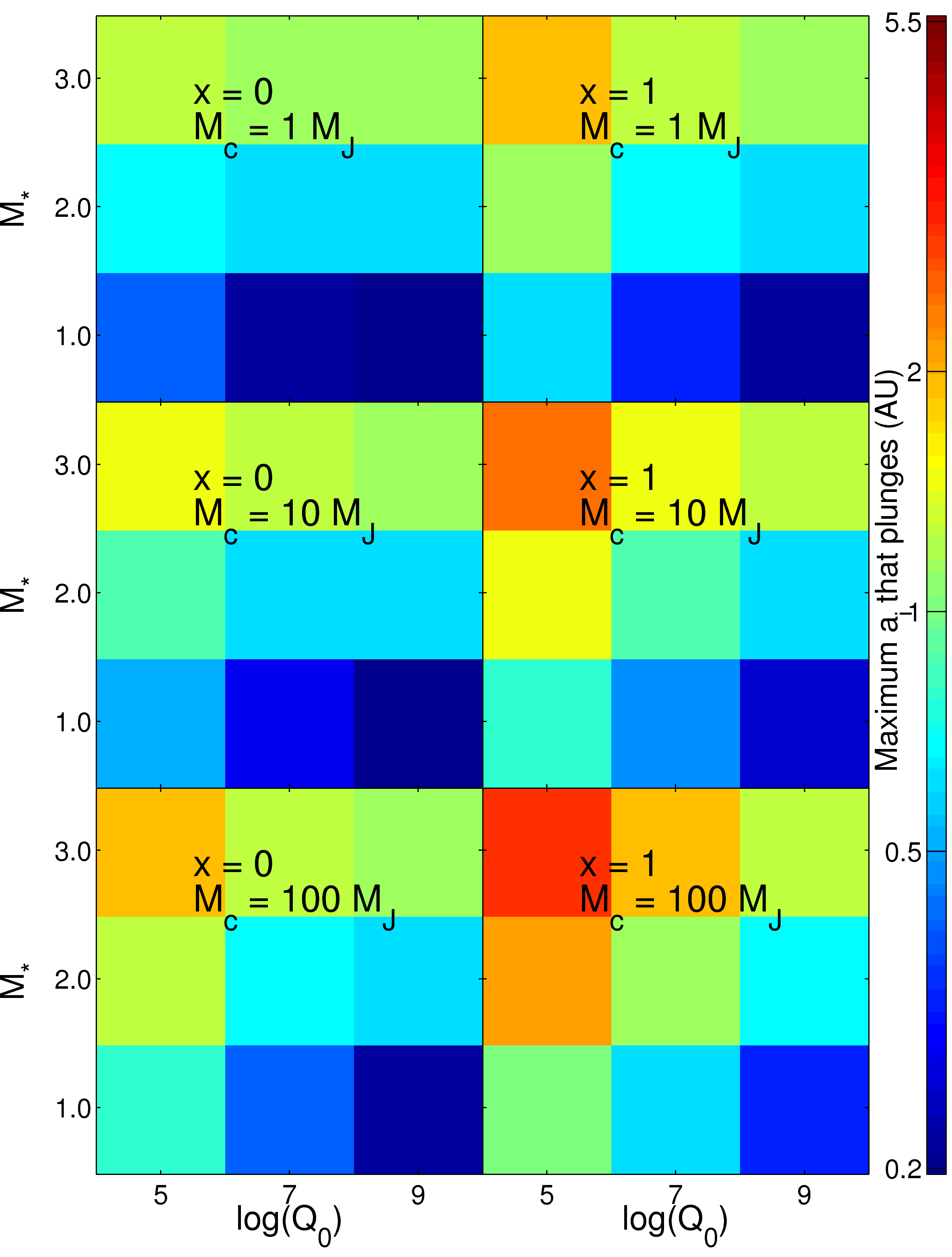}
\end{center}
\caption{Maximum initial semimajor axis, $a$, that enters a common
  envelope for $1$, $10$ and $100$ $M_{\rm J}$ companions with $1$,
  $2$ and $3$ $M_\odot$ primary stars.  The left column presents
  results for $x = 0$ [$Q'_\star = Q_0$] while the right presents
  results for $x = 1$ [$Q'_\star = Q_0\times(\Pi / \Pi_0)$].  The
  color bar is in astronomical units.  The evolutionary models of the
  primary star were computed with a Reimers mass-loss parameter of
  $\eta = 0.7$ (left figure) and $\eta = 5$ (right figure).  In
  general, less mass-loss leads to tidal engulfment at a slightly
  larger semimajor axis.
\label{amax_other_etas}}
\end{figure}

In order to estimate the effects of mass-loss,
Fig.~\ref{amax_other_etas} shows $a_{\rm i,max}$ for $\eta=0.7$ (left
figure) and $\eta=5$ (right figure).  In general, less mass loss leads
to tidal engulfment at slightly larger distances.

By coupling stellar evolution models with various tidal theories, we
determined the maximum separation at which a companion might be
tidally captured.  Companions slightly exterior to $a_{\rm i,max}$ are
never engulfed by the envelope of the primary but their orbital
dynamics are still subject to the effects of mass loss and tidal
torques.  By continuing to follow the orbital evolution, we can
predict the minimum final separation for each binary configuration.
The minimum of this set is the minimum separation exterior to which
one would expect to find planetary companions around white dwarfs.

In the following two sections, we determine the maximum separation at
which a companion can survive CE evolution.  Interior to this
separation, we would expect to find companions to white dwarfs.  Taken
together, the minimum and maximum form a separation gap between which
we expect an absence of planetary and brown dwarf companions around white dwarfs.

\subsection{What About the Solar System?}
\label{ssec:SSystem}
For the solar system, tides and mass-loss compete and determine
whether the inner planets plunge into the Sun or evade tidal
engulfment \citep{Rybicki:2001fk, Sackmann:1993lr}.  Under the
assumption $f=1$, Venus plunges into the Sun for all of our
1~$M_\odot$ evolutionary models.  Earth however, evades tidal
engulfment with the orbit expanding for all of our evolutionary
models, consistent with the results of \citet{Rasio:1996yq}.  In
addition, none of the planets beyond Earth are swallowed.

\section{Common Envelope Evolution}
\label{sec:CEevolve}
Upon tidal engulfment, the companion enters a common envelope with the
primary star \citep{Paczynski:1976fj, Iben:1993kx, Nordhaus:2006oq}.
The velocity difference between the orbital motion of the companion
and the common envelope generates drag.  The resulting loss of orbital
energy leads to rapid in-spiral on week- to month-timescales
\citep{Nordhaus:2006oq}.  The orbital energy released as the companion
in-spirals can be used to overcome the binding energy of the envelope.
If sufficient energy is released from the orbit during in-spiral, the
companion can eject the envelope and survive the CE.  This is
expressed as the following:
\begin{equation}
E_{\rm bind} = -\alpha\Delta E_{\rm orb} \label{CE},
\end{equation}
where $E_{\rm bind}$ is the binding energy of the envelope, $\alpha$
is the fraction of orbital energy that goes toward ejecting the
envelope and $\Delta E_{\rm orb}$ is the change in orbital energy of
the companion.

The change in orbital energy of the companion is given as $E_{{\rm
    orb},R_\star} - E_{{\rm orb},a}$, where $E_{{\rm orb},R_\star}$ is
the orbital energy at tidal engulfment ($a = R_\star$) and $E_{{\rm
    orb}, a}$ is the orbital energy at semimajor axis $a$ inside the
CE (i.e. $a<R_\star$).  The orbital energy at tidal engulfment is
given by the sum of gravitational and potential energies, namely:
\begin{equation}
E_{{\rm orb},R_\star} = - \frac{GM_\star\left[R_\star\right]M_c}{2R_\star}\nonumber
\end{equation}
where $M_\star[R_\star]$ is the total mass of the primary.  The
orbital energy at $a<R_\star$ is given by the kinetic energy ($K_a$) plus the
gravitational potential energy ($U_a$):
\begin{eqnarray}
E_{{\rm orb},a} & = &K_{a}+ U_{a}\\
 & = &  \frac{GM_\star\left[a\right]M_c}{2a} - \frac{GM_\star\left[R_\star\right]M_c}{R_\star} -  \int^{R_\star}_{a}\frac{GM_\star\left[r\right]M_c}{r^2}dr \, , \nonumber
\end{eqnarray}
where the integral term is the work against gravity required to move
the companion from $a$ to$R_\star$.  The total change in orbital
energy is then given as:
\begin{eqnarray}
\Delta E_{\rm orb} & = & E_{{\rm orb}, a}\label{Delta_E_orb} - E_{{\rm orb},R_\star} \\
&=& \frac{G M_\star[a]M_c}{2a} - \frac{G M_\star[R_\star]M_c}{2R_\star} -\int_a^{R_\star}\frac{GM_\star[r]M_c}{r^2}dr \, .\nonumber
\end{eqnarray}

The companion in-spirals until it is either tidally disrupted or
supplies enough orbital energy such that $-\alpha \Delta E_{\rm orb} =
E_{\rm bind}$ \citep{Nordhaus:2006oq}.  The binding energy of the
stellar envelope is computed by summing the gravitational and thermal
energies, at each evolutionary timestep.
We require that the companion supplies enough orbital energy to
overcome the binding energy of the entire envelope.  If the companion
only ejected the mass exterior to its orbit, the interior mass would
expand to re-engulf the companion and restart in-spiral.  Note that
Eq.~(\ref{Delta_E_orb}) depends on the stellar structure at the onset
of tidal engulfment.

\section{Period Gaps for Planets and Brown Dwarfs Around White Dwarfs}
\label{sec:PGaps}
To calculate a minimum period gap expected for a given binary system, we
assume $\alpha = 1$.  This gives an upper bound on the orbital radius
at which we would expect to find companions which have survived a
common envelope phase (see Fig.~\ref{thegap}).  In \S4, we determined
the time at which a companion of mass $M_c$ is engulfed by the giant
star.  Coupled with the structure of the star at the time of tidal
engulfment, we can determine which semimajor axis $a$ satisfies
Eq.~(\ref{CE}).  If the companion avoids tidal disruption, then it has
successfully ejected the envelope and survived the CE phase.

The tidal shredding radius can be estimated by balancing the
differential gravitational force across the companion with its self
gravity.  This yields a tidal shredding radius given by $a_s \simeq
R_c \sqrt[3]{2M / M_c}$ where $R_c$ is the radius of the companion.
For a $1M_{\rm J}$ companion around a proto-white dwarf core, $a_s
\sim 7.4\times 10^{10}$~cm.  For a 10~$M_{\rm J}$ companion\footnote{$M_{\rm J}$ is
  the mass of Jupiter.} $a_s\gtrsim 3\times 10^{10}$~cm with the actual values dependent on the degenerate core mass during the CEP.  If the companion unbinds the envelope exterior to $a_s$, then we say it has
survived common envelope evolution.  Note that the factor of $2M/M_c$ in the expression for $a_s$ neglects the synchronous rotation of the companion and its finite Love number.  Including these effects leads to a slightly larger tidal shredding radius, implying that slightly more massive companions are required to unbind the CE and avoid tidal disruption.  

Since the stellar mass function is heavily weighted towards lower
masses, we limit ourselves to a 1~$M_\odot$ progenitor.  More massive
primaries extend to larger radii (see Fig.~\ref{models}) and swallow
companions at farther distances, leading to wider period gaps.  We
consider two companions: a 1~$M_{\rm J}$ planet and a 10~$M_{\rm J}$ brown dwarf.

For each evolutionary model and binary configuration, we calculate the
minimum and maximum bounds of the separation gap.  Our results for
$\eta = 0.7$, $1$ and $5$, $x=0$ and $1$ and $Q_0 = 10^6$ and $f=1$
are summarized in Table \ref{table2}.  The
maximum of $a_{\rm min}$ and minimum of $a_{\rm max}$ yield the
minimum gap expected for a $M_{\star} = 1 M_\odot$, $M_{\rm c}=1
M_{\rm J}$ system.  Note that for pedagogical purposes we include the results for mass-loss only (no tides) in Table \ref{table2}.  However, when determining minimum period gaps, we only consider systems in which tides are acting.   It should be stressed that even in the absence of tides, a minimum period gap exists.  The functional dependence of tides on separation (Eq.~2) largely acts to change the location of the outer boundary of the gap.  Depending on tidal prescription, this shifts the outer boundary of the by a factor of $\sim$2-3 at most (Table \ref{table2}).  

From Table \ref{table2}, we see that there should
be a paucity of Jupiter-mass companions with periods
$\lesssim$270~days around white dwarfs.  Additionally from Table
\ref{table2}, we see that there should be a paucity of 10~$M_{\rm J}$
companions with periods between 0.1~days (0.003~AU) and 380~days
(0.75~AU).  This is consistent with the tentative detection of a
$\sim$2~$M_{\rm J}$ planet in a $\gtrsim$4~year ($\gtrsim$2.75~AU)
orbit around the white dwarf GD 66 \citep{Mullally:2008fk,
  Mullally:2009uq}.  Future surveys for low-mass companions around
white dwarfs will have the ability to confirm or refute our prediction
of a gap.  Several efforts are either recently completed or are
currently underway \citep{Farihi:2006vn, Tremblay:2007fr,
  Hoard:2007ys, Farihi:2008rt, Farihi:2009zr}.  Once the samples are
sufficiently large, observational identification of period gaps could
help to constrain aspects of mass-loss and tidal theories.

For each $M_\star = 1 M_\odot$ model, we can calculate the minimum
mass companion able to unbind the envelope and survive a CE phase.  This occurs when the binding energy of envelope is at a minimum.  Note that, in multiple planet systems, several close companions may
incur a CE phase.  In conjunction, multiple lower-mass companions
could potentially supply the same energy as a single larger mass
object.  Although we do not investigate such a scenario, multiple
companions may successfully expel the envelope such that all or a
subset survive.  For the $\eta=0.7$ model, the binding energy of the envelope is minimized when past the peak radius on the AGB.  An 8.4~$M_{\rm J}$ mass companion is sufficient to unbind the envelope and survive a CE phase.  For the $\eta=1$
model, the envelope binding energy is minimized at the peak of the RGB.  For this
system, a 9.8~$M_{\rm J}$ companion is sufficient to survive a CE phase.  The $\eta=5$ model does not achieve a sufficient
core mass to undergo a helium flash.  As such, its maximal radial
extent is less than the $\eta=0.7$ and $\eta=1$ models while the
envelope binding energy is reduced.  In this case, a $\sim$7.8~$M_{\rm J}$
mass companion is sufficient to survive a CE phase.  For all of our models, most  companions more massive than $\sim$10 $M_{\rm J}$ will supply sufficient energy to unbind the envelope and survive the CEP at greater orbital separations.\footnote{Note that even companions whose masses are several tens of Jupiter masses will still be within a few solar radii after surviving inspiral.}  


\begin{table*}
 \centering
 \begin{minipage}{140mm}
  \begin{tabular}{@{}llllllllllllr@{}}
  \hline
  &  &  $M_\star$ & $M_c$ & $Q_0$ & $x$  & $a_{\rm i, crit}$& $M_{\rm core,1}$& $M_{\rm core,2}$ & $a_{\rm min}$\tablenotemark{a}    & $a_{\rm max}$ & $P_{\rm min}$ & $P_{\rm max}$\\
&&   ($M_\odot$) & ($M_{\rm J}$) & & & (AU) &($M_\odot$)& ($M_\odot$)& (AU) & (AU) & (days) & (days)\\
\hline
     &$\eta = 0.7$& 1.0 & 1.0 & -- & --&1.90 &-- & 0.60 & -- & 2.36 & -- & 1700\\ 
       &$\eta = 1$& 1.0 & 1.0 & -- & -- &1.64 &--& 0.56& -- & 1.98 & -- & 1350\\
     &$\eta = 5$\tablenotemark{b}& 1.0 & 1.0 & -- & -- &0.64 &-- &0.38 &-- & 0.79 & -- & 414\\
   $f=1$       &$\eta = 0.7$& 1.0 & 10.0 & -- & --&2.53 &-- & 0.60& -- & 2.93 & -- & 2350\\ 
        &$\eta = 1$& 1.0 & 10.0 & -- & -- &2.18 &0.45&0.56 & $2.7\times10^{-3}$ & 2.51 & 0.08 & 1930\\
        &$\eta = 5^b$& 1.0 & 10.0 & -- & -- &0.86 &-- & 0.38&-- & 1.09 & -- & 670\\
     \hline
&$\eta = 0.7$& 1.0 & 1.0 & $10^6$ & 0&0.75 &  --& 0.60& -- & 1.20 & -- & 616\\
&$\eta = 0.7$& 1.0 & 1.0 & $10^6$ & 1 &1.10 &--& 0.60&-- & 1.37 & -- & 751\\
 &$\eta = 1$& 1.0 & 1.0 & $10^6$ & 0 &0.67 &--& 0.56&-- & 1.15 & -- & 598\\
  &$\eta = 1$&  1.0 & 1.0 & $10^6$ & 1 &0.97 &--& 0.56&-- & 1.30 & -- & 719\\
  &$\eta = 5$\tablenotemark{b}& 1.0 & 1.0 & $10^6$ & 0&0.29 &-- & 0.38& -- & 0.60 & -- & 274\\
    &$\eta = 5$\tablenotemark{b}& 1.0 & 1.0 & $10^6$ & 1&0.47 &-- & 0.38& -- & 0.87 & -- & 478\\
$Q'_\star \propto \Pi^x$&$\eta = 0.7$& 1.0 & 10.0 & $10^6$ & 0& 0.93& 0.60&0.60 &$2.7\times10^{-3}$ & 1.30 & 0.07 & 695\\
&$\eta = 0.7$& 1.0 & 10.0 & $10^6$ & 1 & 1.46&0.60& 0.60&$3.0\times10^{-3}$ & 1.65 & 0.08 & 993\\
 &$\eta = 1$& 1.0 & 10.0 & $10^6$ & 0 &0.79 &--& 0.56&-- & 1.32 & -- & 736\\
  &$\eta = 1$& 1.0 & 10.0 & $10^6$ & 1 &1.30 &--& 0.56&-- & 1.62 & -- & 1000\\
     &$\eta = 5$\tablenotemark{b}& 1.0 & 10.0 & $10^6$ & 0&0.41 &-- &0.38 & -- & 0.75 & -- & 383\\
      &$\eta = 5$\tablenotemark{b}& 1.0 & 10.0 & $10^6$ & 1&0.63 &-- & 0.38& -- & 1.16 & -- & 736\\
  \hline
   &$\eta = 0.7$& 1.0 & 1.0 & -- & --&0.73 &-- & 0.60& -- & 1.21 & -- & 624\\
   &$\eta = 1$& 1.0 & 1.0 & -- & -- &0.65 &--& 0.56& -- & 1.15 & -- & 598\\
 &$\eta = 5$\tablenotemark{b}& 1.0 & 1.0 & -- & -- &0.22 &-- & 0.38& -- & 0.57 & -- & 253\\
 No tides&$\eta = 0.7$& 1.0 & 10.0 & -- & --&0.73 &-- & 0.60& -- & 1.21 & -- & 624\\
 &$\eta = 1$& 1.0 & 10.0 & -- & -- & 0.65&0.45& 0.56& $2.7\times10^{-3}$ & 1.15 & 0.08 & 598\\
         &$\eta = 5$\tablenotemark{b}& 1.0 & 10.0 & -- & -- &0.22 &-- &0.38 & -- & 0.57 & -- & 253\\
     \hline
       &$\eta = 0.7$& 1.0 & 8.4\tablenotemark{c}  & -- & --&-- &0.60 & --& $2.5\times10^{-3}$& -- & 0.06 & --\\
   &$\eta = 1$& 1.0 & 9.8\tablenotemark{c} & -- & -- &-- &0.45& --& $2.2\times10^{-3}$ & -- & 0.06 & --\\
&$\eta = 5$& 1.0 & 7.8\tablenotemark{c} & -- & -- &-- &0.38& --&$2.2\times10^{-3}$ & -- & 0.06 & --\\
$E_{\rm bind,min}$ &$\eta = 0.7$& 1.0 & 10.0 & -- & --&-- &0.60 & --&$3.0\times10^{-3}$ & -- & 0.08 & --\\
 &$\eta = 1$& 1.0 & 10.0 & -- & -- & --&0.45& --& $2.7\times10^{-3}$ & -- & 0.08 & --\\
        &$\eta = 5$& 1.0 & 10.0 & -- & -- &-- &0.38&-- & $2.9\times10^{-3}$ & -- & 0.09& --\\
        \hline
  \end{tabular}
$^a$~{Companions which do not survive a common envelope phase are denoted by a $-$.}\\
$^b$~{The $\eta=5$ model does not achieve a sufficient core mass to undergo a helium flash.}\\
$^c$~{The minimum mass companion required to unbind the envelope and survive a CEP.} 
\end{minipage}
 \caption{Period gaps for binary systems with different tidal formalisms ($Q'_\star \propto \Pi^x$; $f=1$; no tides).  The column quantities
   from left to right are Reimers mass-loss parameter $\eta$, $M_\star$ the mass of the primary star, $M_c$ the mass
   of the companion, $Q_0$ [Eq. (\ref{eq:Jeremy}], $x$ values such that $Q'_\star = Q_0 (\Pi /
   \Pi_0)^x$ [Eq. \ref{eq:Jeremy}], $a_{\rm i,crit}$, the critical initial radius below/above which the companion is/isn't engulfed (see Fig. \ref{thegap}), $M_{\rm core,1}$, the core mass for $a_{\rm min}$, $M_{\rm core,2}$, the core mass for $a_{\rm
     max}$, $a_{\rm min}$ the minimum separation (from CE evolution), $a_{\rm max}$ the maximum separation
   (from orbital evolution), $P_{\rm min}$ the minimum period (from CE evolution),
   $P_{\rm max}$ the maximum period (from orbital evolution).  Companions that do not
   survive common envelope evolution are denoted by $-$.  Note that the time of engulfment for the above cases does not necessarily coincide with the minimum binding energy of the envelope during its evolution.  Therefore, we include results for companions entering the giant star at the time of minimum envelope binding energy ($E_{\rm bind, min}$). For
   graphical representations of this table see Fig. \ref{table_gap}.
\label{table2}}
\end{table*}

\begin{figure}
\begin{center}
\includegraphics[width=8.5cm,angle=0,clip=true]{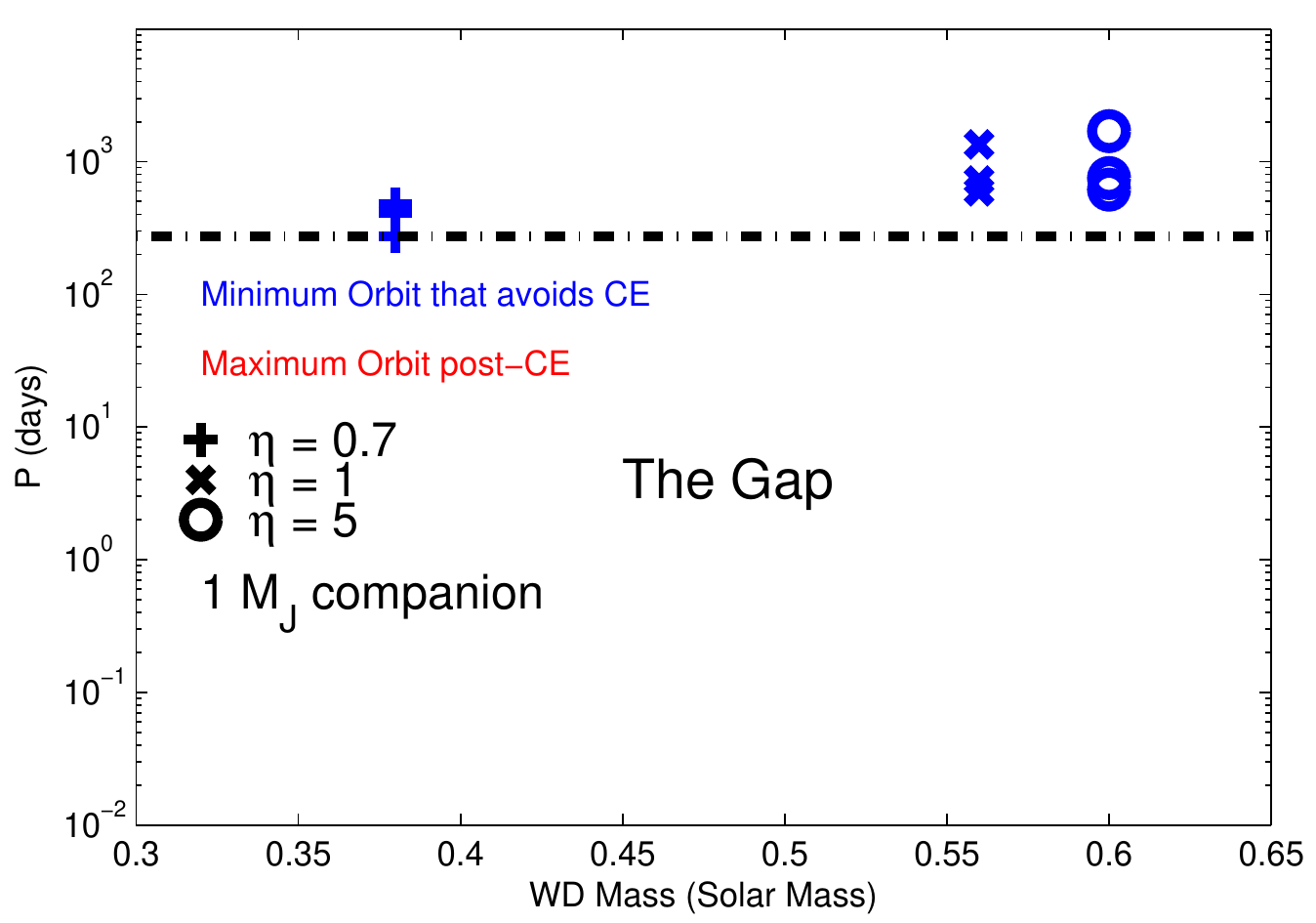}
\includegraphics[width=8.5cm,angle=0,clip=true]{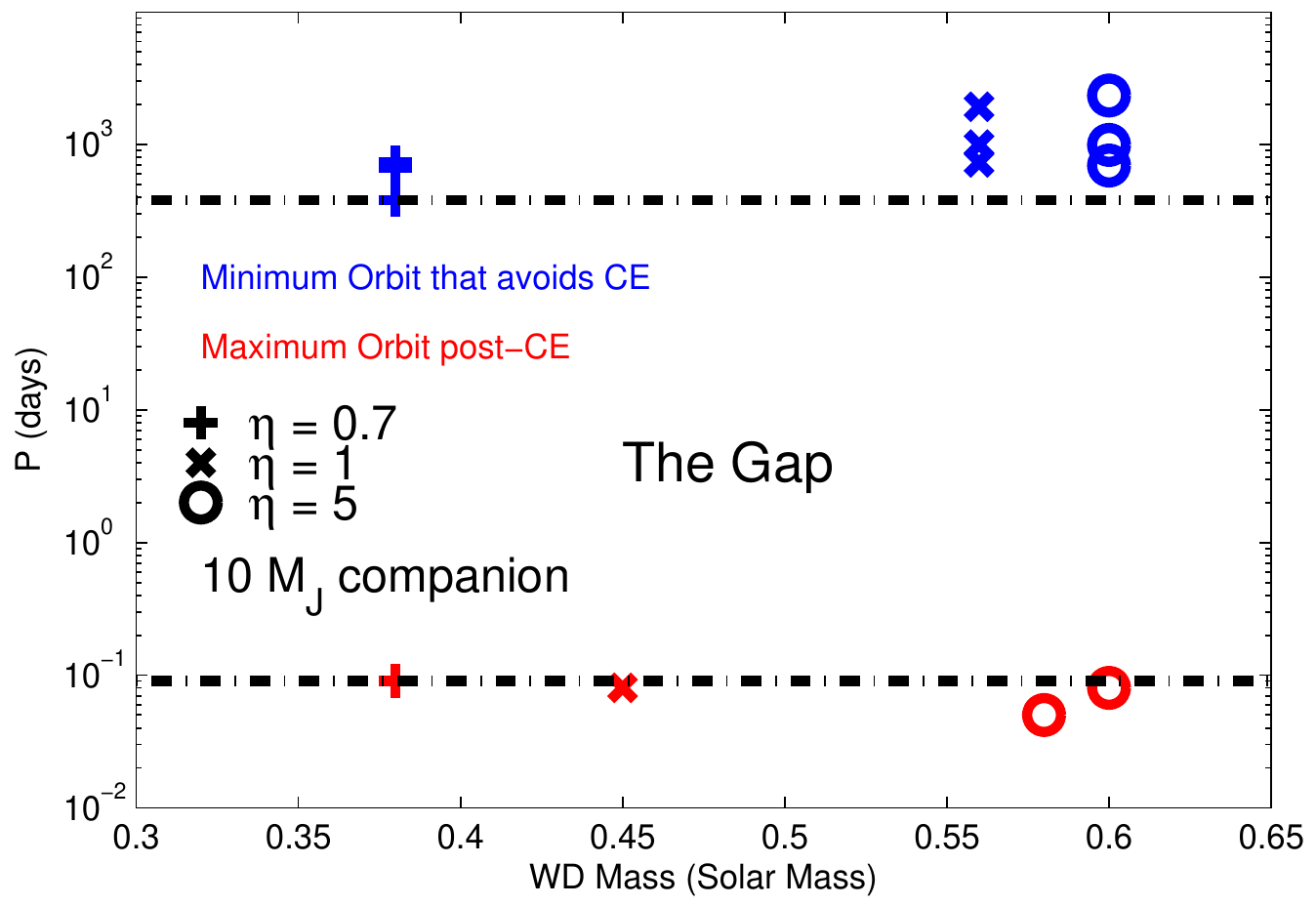}
\caption{The predicted period gaps for a 1 $M_\odot$ progenitor with 1
  $M_{\rm J}$ (top) and 10 $M_{\rm J}$ (bottom) companions.  The symbols represent
  different Remiers $\eta$ values for the various tidal prescriptions
  listed in Table \ref{table2}.  For the 1 $M_{\rm J}$ system, no companion
  survives CE evolution.  Thus, we predict a paucity of 1 $M_{\rm J}$
  companions with periods $\lesssim$270 days.  For the 10 $M_\odot$
  system, several companions survive CE evolution and are located in
  short-period orbits.  The predicted period gap occurs between
  $\sim$0.1 and 380 days.
\label{table_gap}}
\end{center}
\end{figure}

\section{WD Transit Detectability and GALEX}
\label{sec:WD_detect}
If a massive planetary companion or a brown dwarf is engulfed and
survives the common envelope phase, it could be left in a close,
short-period orbit around the resultant WD.  In such a scenario, if
the orbit were oriented edge-on, it is conceivable that an entirely
new regime of exoplanetary transiting observations would allow for
detection of the companion.  In contrast to the observations pioneered
by \citet{Henry:2000lr} and \citet{Charbonneau:2000fk}\footnote{See
  http://exoplanet.eu for an up-to-date census of the now more than 60
  known transiting planets, and of all other known exoplanets.}, where
transiting Jupiter-sized planets cause dips of $\sim$1\% in the light
curves of main sequence solar type stars, a massive Jupiter transiting
a WD could cause a total eclipse (a 100\% dip!).  How likely are such
detections?  In this context, rather than the oft-quoted a priori
probability of $R_*/a$ that an exoplanet's orbit would be oriented so
as to transit a main sequence star, the probability of a total eclipse
in a system where $R_p > R_*$ is
\begin{equation}
p_{\rm orbit} \sim \frac{R_p}{a} \sim 0.1 \left(\frac{R_p/R_J}{a / R_\odot}\right) \, ,
\label{eq:Porbit}
\end{equation}
where $R_J$ is Jupiter's radius, indicating that roughly one out of
every $\sim$10 such post-CE systems might have an orbit that is
oriented so that the companion eclipses the WD once per orbit.  The
duration of the eclipses will be of order
\begin{equation}
\delta_{\rm eclipse} \sim \frac{R_p}{2\pi a} \sim  2\times 10^{-2} \left( \frac{R_p/R_J}{a/R_\odot} \right)
\label{eq:reclipse}
\end{equation}
times the length of the year ($P$), or
\begin{equation}
\Delta t_{\rm eclipse} = \delta_{\rm eclipse} \times P \sim (4 {\rm~min}) (R_p/R_J) \sqrt{ \frac{a / R_\odot}{M / (0.6 M_\odot)} } \, .
\label{eq:teclipse}
\end{equation}
At any given time, the fraction of WDs being eclipsed by companions is of order
\begin{eqnarray}
f_{\rm eclipse} &\sim& f_{\rm WD,p} \times \tilde{p}_{\rm orbit} \times \tilde{\delta}_{\rm eclipse} \\
\nonumber &\sim& 2\times 10^{-3} f_{\rm WD,p} \left( \frac{\tilde{R}_p/R_J}{\tilde{a}/R_\odot} \right)^2 \, ,
\label{eq:feclipse}
\end{eqnarray}
where $f_{\rm WD,p}$ is the fraction of WDs with short-period
companions around them and $\tilde{p}_{\rm orbit}$, $\tilde{\delta}_{\rm
  eclipse}$, $\tilde{R}_p$, and $\tilde{a}$ are typical values of
$p_{\rm orbit}$, $\delta_{\rm eclipse}$, $R_p$, and $a$, respectively.  The
Galaxy Evolution Explorer ({\it GALEX}) is an ultraviolet space
telescope capable of observing many hot WDs.  Since {\it GALEX}
time-stamps photons, it should be relatively easy to search the {\it
  GALEX} archive for transits.  In order to detect an eclipsing
system, a WD must be observed for a duration of at least $\Delta
t_{\rm eclipse}$, or at two different epochs separated by at least
$\Delta t_{\rm eclipse}$ such that it ``winks out'' and then
reappears.  Depending on the number of WDs that {\it GALEX} has
observed (and on the duration of observations), a search of the {\it
  GALEX} archive might either detect some WD eclipsing binary systems
-- contingent on the distribution of radii and orbital semimajor axes
of massive planets and brown dwarfs around WDs -- or place constraints
on the population of short-period companions to WDs.

\subsection{Progenitors of Cataclysmic Variables}

Via loss of orbital angular momentum, the short period systems that survive a CEP may be brought into Roche contact with the WD.  Such systems may have been detected as cataclysmic variables (CV) but not recognized as post-planetary systems.  If the donor is a massive planet with a rocky core, it may be possible for the core to survive the CV phase.  This would result in a WD+rocky core short period binary.  

\section{Conclusions}
\label{sec:conc}
By utilizing stellar evolution models from the ZAMS through the
post-MS, we have followed the orbital dynamics of binary systems in
which the companion is a 1~$M_{\rm J}$ planet, 10~$M_{\rm J}$ brown
dwarf, or 100~$M_{\rm J}$ low-mass MS star.  Our evolutionary models
incorporate a range of mass-loss rates and stellar masses.
Dynamically, the orbital evolution is subject to mass-loss (which acts
to increase the separation) and tidal torques (which act to decrease
the separation).

We employ two commonly used tidal prescriptions to investigate the
differences during the post-MS.  In particular, we find the following:

\begin{itemize}
\item The tidal theory of \citet{Goldreich:1963nr} under the assumption that $Q'_\star\sim10^5-10^9$ leads to relatively weak tides, often capturing companions whose initial orbits are within $\sim0.6\times R_{\rm max}$, where $R_{\rm max}$ is the maximum stellar radius for an evolutionary model.  Companions with initial orbits larger than $0.6 R_{\max}$ but smaller than $R_{\max}$ can escape capture since their orbits expand due to mass loss of the primaries.  
\item The tidal theory of \citet{Zahn:1966jk}, under the assumption
  that $f=1$ \citep{Verbunt:1995rt}, leads to comparatively strong tidal torques,
  often capturing companions within $\sim$$(2-3)\times R_{\rm max}$.  Nevertheless, even this is less than the values of $\sim$$5-7\times R_{\rm max}$ that have been quoted in the literature \citep{Soker:1996fk,Debes:2002kx,Moe:2006fc}.
\item The ratio $a_{\rm i,max} / R_\star$, where $a_{\rm i,max}$ is
  the maximum radius that is tidally engulfed, is not constant.
\item Future determination of the observational period gap for
  low-mass companions around WDs should help place constraints on the
  actual tidal theory acting in the post-MS.
\end{itemize}

For each tidal theory, we determined the maximum separation for which
companions might be tidally engulfed (i.e. plunge into the primary
star).  These results serve as initial conditions for the onset of the
common envelope phase for low-mass companions.  Previous population
synthesis predictions for post-AGB stars and PNe can be refined by
incorporating the methods outlined in this paper.

For companions that incur a CE, under the assumption of maximum
orbital energy deposition to the common envelope, we determined the
maximum orbital radius at which a companion survives the interaction.
By following the orbital evolution of the closest companion that
evades tidal engulfment, we predict a period gap for each binary
system.

For a binary system consisting of a 1~$M_\odot$ primary with a
1~$M_{\rm J}$ companion, we predict a paucity of Jupiter-mass
companions with period below $\sim$270~days.  For a 1~$M_\odot$
primary with a 10~$M_{\rm J}$ companion, the gap occurs between
$\sim$0.1 and $\sim$380~days corresponding to $\sim$0.003-0.75~AU.
Note that our estimated gaps are conservative and are obtained by
finding the minimum gap that might be expected for a range of
mass-loss rates and a range of assumptions about tidal dissipation.
It is unlikely that the true gaps would be narrower than the ranges
quoted above, but they easily could be wider.  As our knowledge of
stellar evolution and tidal dissipation improves, so will our
estimates of the ranges for these gaps.  Finally, we note that the
results of surveys searching for low mass companions to white dwarfs
might help to constrain theories of both stellar evolution and tides.

\section*{Acknowledgments}
We thank Brad Hansen, Frank Verbunt, Alex Wolszczan, Orsola De Marco,
Eric Blackman, Jay Farihi, David Hogg, David Schiminovich, Claire Lackner, Fergal
Mullally, Sarah Wellons, Rich Gott, and Kurtis Williams, for
thoughtful discussions and comments.

Support for HPC storage and resources was provided by the TIGRESS High
Performance Computing and Visualization Center at Princeton
University.  JG acknowledges partial support via NASA grant
AST-0707373.  AB and DS acknowledge support by NASA grant NNX07AG80G
and under JPL/Spitzer Agreements 1328092, 1348668, and 1312647.  This
research was supported in part by the National Science Foundation
under Grant No. PHY05-51164.

\bibliography{general}

\begin{thebibliography}{}

\bibitem[\protect\astroncite{{Arras} and {Socrates}}{2009}]{Arras:2009rt}
{Arras}, P. and {Socrates}, A.: 2009,
\newblock {\em ArXiv e-prints}

\bibitem[\protect\astroncite{{Balick} and {Frank}}{2002}]{Balick:2002yf}
{Balick}, B. and {Frank}, A.: 2002,
\newblock {\em \araa} {\bf 40}, 439

\bibitem[\protect\astroncite{{Barnes} et~al.}{2008}]{Barnes:2008lr}
{Barnes}, R., {Raymond}, S.~N., {Jackson}, B., and {Greenberg}, R.: 2008,
\newblock {\em Astrobiology} {\bf 8}, 557

\bibitem[\protect\astroncite{{Bear} and {Soker}}{2009}]{Bear:2009yq}
{Bear}, E. and {Soker}, N.: 2009,
\newblock {\em ArXiv e-prints}

\bibitem[\protect\astroncite{{Bujarrabal} et~al.}{2001}]{Bujarrabal:2001bs}
{Bujarrabal}, V., {Castro-Carrizo}, A., {Alcolea}, J., and {S{\'a}nchez
  Contreras}, C.: 2001,
\newblock {\em \aap} {\bf 377}, 868

\bibitem[\protect\astroncite{{Burrows} et~al.}{2001}]{Burrows:2001fk}
{Burrows}, A., {Hubbard}, W.~B., {Lunine}, J.~I., and {Liebert}, J.: 2001,
\newblock {\em Reviews of Modern Physics} {\bf 73}, 719

\bibitem[\protect\astroncite{{Burrows} et~al.}{1993}]{Burrows:1993zv}
{Burrows}, A., {Hubbard}, W.~B., {Saumon}, D., and {Lunine}, J.~I.: 1993,
\newblock {\em \apj} {\bf 406}, 158

\bibitem[\protect\astroncite{{Burrows} et~al.}{1997}]{Burrows:1997lr}
{Burrows}, A., {Marley}, M., {Hubbard}, W.~B., {Lunine}, J.~I., {Guillot}, T.,
  {Saumon}, D., {Freedman}, R., {Sudarsky}, D., and {Sharp}, C.: 1997,
\newblock {\em \apj} {\bf 491}, 856

\bibitem[\protect\astroncite{{Carlberg} et~al.}{2009}]{Carlberg:2009uq}
{Carlberg}, J.~K., {Majewski}, S.~R., and {Arras}, P.: 2009,
\newblock {\em \apj} {\bf 700}, 832

\bibitem[\protect\astroncite{{Charbonneau} et~al.}{2000}]{Charbonneau:2000fk}
{Charbonneau}, D., {Brown}, T.~M., {Latham}, D.~W., and {Mayor}, M.: 2000,
\newblock {\em \apjl} {\bf 529}, L45

\bibitem[\protect\astroncite{{Chesneau} et~al.}{2009}]{Chesneau:2009lr}
{Chesneau}, O., {Clayton}, G.~C., {Lykou}, F., {de Marco}, O., {Hummel}, C.~A.,
  {Kerber}, F., {Lagadec}, E., {Nordhaus}, J., {Zijlstra}, A.~A., and {Evans},
  A.: 2009,
\newblock {\em \aap} {\bf 493}, L17

\bibitem[\protect\astroncite{{de Marco}}{2009}]{de-Marco:2009vl}
{de Marco}, O.: 2009,
\newblock {\em \pasp} {\bf 121}, 316

\bibitem[\protect\astroncite{{De Marco} et~al.}{2008}]{De-Marco:2008nx}
{De Marco}, O., {Hillwig}, T.~C., and {Smith}, A.~J.: 2008,
\newblock {\em \aj} {\bf 136}, 323

\bibitem[\protect\astroncite{{De Marco} and {Moe}}{2005}]{De-Marco:2005xw}
{De Marco}, O. and {Moe}, M.: 2005,
\newblock in R. {Szczerba}, G. {Stasi{\'n}ska}, and S.~K. {Gorny} (eds.), {\em
  Planetary Nebulae as Astronomical Tools}, Vol. 804 of {\em American Institute
  of Physics Conference Series}, pp 169--172

\bibitem[\protect\astroncite{{Debes} and {Sigurdsson}}{2002}]{Debes:2002kx}
{Debes}, J.~H. and {Sigurdsson}, S.: 2002,
\newblock {\em \apj} {\bf 572}, 556

\bibitem[\protect\astroncite{{Farihi}}{2009}]{Farihi:2009zr}
{Farihi}, J.: 2009,
\newblock {\em \mnras} {\bf 398}, 2091

\bibitem[\protect\astroncite{{Farihi} et~al.}{2008}]{Farihi:2008rt}
{Farihi}, J., {Becklin}, E.~E., and {Zuckerman}, B.: 2008,
\newblock {\em \apj} {\bf 681}, 1470

\bibitem[\protect\astroncite{{Farihi} et~al.}{2006}]{Farihi:2006vn}
{Farihi}, J., {Hoard}, D.~W., and {Wachter}, S.: 2006,
\newblock {\em \apj} {\bf 646}, 480

\bibitem[\protect\astroncite{{Ferraz-Mello} et~al.}{2008}]{Ferraz-Mello:2008dk}
{Ferraz-Mello}, S., {Rodr{\'{\i}}guez}, A., and {Hussmann}, H.: 2008,
\newblock {\em Celestial Mechanics and Dynamical Astronomy} {\bf 101}, 171

\bibitem[\protect\astroncite{{Goldreich} and
  {Nicholson}}{1977}]{Goldreich:1977qy}
{Goldreich}, P. and {Nicholson}, P.~D.: 1977,
\newblock {\em Icarus} {\bf 30}, 301

\bibitem[\protect\astroncite{{Goldreich} and {Soter}}{1966}]{Goldreich:1966qv}
{Goldreich}, P. and {Soter}, S.: 1966,
\newblock {\em Icarus} {\bf 5}, 375

\bibitem[\protect\astroncite{{Goldreich}}{1963}]{Goldreich:1963nr}
{Goldreich}, R.: 1963,
\newblock {\em \mnras} {\bf 126}, 257

\bibitem[\protect\astroncite{{Goodman} and {Lackner}}{2009}]{Goodman:2009kx}
{Goodman}, J. and {Lackner}, C.: 2009,
\newblock {\em \apj} {\bf 696}, 2054

\bibitem[\protect\astroncite{{Gray}}{1989}]{Gray:1989qa}
{Gray}, D.~F.: 1989,
\newblock {\em \apj} {\bf 347}, 1021

\bibitem[\protect\astroncite{{Greenberg}}{2009}]{Greenberg:2009lr}
{Greenberg}, R.: 2009,
\newblock {\em \apjl} {\bf 698}, L42

\bibitem[\protect\astroncite{{Gu} and {Ogilvie}}{2009}]{Gu:2009yq}
{Gu}, P. and {Ogilvie}, G.~I.: 2009,
\newblock {\em \mnras} {\bf 395}, 422

\bibitem[\protect\astroncite{{Henry} et~al.}{2000}]{Henry:2000lr}
{Henry}, G.~W., {Marcy}, G.~W., {Butler}, R.~P., and {Vogt}, S.~S.: 2000,
\newblock {\em \apjl} {\bf 529}, L41

\bibitem[\protect\astroncite{{Hoard} et~al.}{2007}]{Hoard:2007ys}
{Hoard}, D.~W., {Wachter}, S., {Sturch}, L.~K., {Widhalm}, A.~M., {Weiler},
  K.~P., {Pretorius}, M.~L., {Wellhouse}, J.~W., and {Gibiansky}, M.: 2007,
\newblock {\em \aj} {\bf 134}, 26

\bibitem[\protect\astroncite{{Iben} and {Livio}}{1993}]{Iben:1993kx}
{Iben}, Jr., I. and {Livio}, M.: 1993,
\newblock {\em \pasp} {\bf 105}, 1373

\bibitem[\protect\astroncite{{Ibgui} and {Burrows}}{2009}]{Ibgui:2009jk}
{Ibgui}, L. and {Burrows}, A.: 2009,
\newblock {\em \apj} {\bf 700}, 1921

\bibitem[\protect\astroncite{{Ibgui} et~al.}{2010}]{Ibgui:2010lr}
{Ibgui}, L., {Burrows}, A., and {Spiegel}, D.~S.: 2010,
\newblock {\em \apj} {\bf 713}, 751

\bibitem[\protect\astroncite{{Ibgui} et~al.}{2009}]{Ibgui:2009fk}
{Ibgui}, L., {Spiegel}, D.~S., and {Burrows}, A.: 2009,
\newblock {\em ArXiv e-prints}

\bibitem[\protect\astroncite{{Jackson} et~al.}{2009}]{Jackson:2009bq}
{Jackson}, B., {Barnes}, R., and {Greenberg}, R.: 2009,
\newblock {\em \apj} {\bf 698}, 1357

\bibitem[\protect\astroncite{{Jackson} et~al.}{2008}]{Jackson:2008kl}
{Jackson}, B., {Greenberg}, R., and {Barnes}, R.: 2008,
\newblock {\em \apj} {\bf 678}, 1396

\bibitem[\protect\astroncite{{Jacoby}}{1980}]{Jacoby:1980kc}
{Jacoby}, G.~H.: 1980,
\newblock {\em \apjs} {\bf 42}, 1

\bibitem[\protect\astroncite{{Kaula}}{1968}]{Kaula:1968fk}
{Kaula}, W.~M.: 1968,
\newblock {\em {An introduction to planetary physics - The terrestrial
  planets}}

\bibitem[\protect\astroncite{{Kilic} et~al.}{2010}]{Kilic:2010qy}
{Kilic}, M., {Brown}, W.~R., and {McLeod}, B.: 2010,
\newblock {\em \apj} {\bf 708}, 411

\bibitem[\protect\astroncite{{Levrard} et~al.}{2009}]{Levrard:2009qy}
{Levrard}, B., {Winisdoerffer}, C., and {Chabrier}, G.: 2009,
\newblock {\em \apjl} {\bf 692}, L9

\bibitem[\protect\astroncite{{Livio} and {Soker}}{1984}]{Livio:1984fk}
{Livio}, M. and {Soker}, N.: 1984,
\newblock {\em \mnras} {\bf 208}, 763

\bibitem[\protect\astroncite{{Massarotti} et~al.}{2008}]{Massarotti:2008rw}
{Massarotti}, A., {Latham}, D.~W., {Stefanik}, R.~P., and {Fogel}, J.: 2008,
\newblock {\em \aj} {\bf 135}, 209

\bibitem[\protect\astroncite{{Maxted} et~al.}{2006}]{Maxted:2006fj}
{Maxted}, P.~F.~L., {Napiwotzki}, R., {Dobbie}, P.~D., and {Burleigh}, M.~R.:
  2006,
\newblock {\em \nat} {\bf 442}, 543

\bibitem[\protect\astroncite{{Miller} et~al.}{2009}]{Miller:2009lr}
{Miller}, N., {Fortney}, J.~J., and {Jackson}, B.: 2009,
\newblock {\em \apj} {\bf 702}, 1413

\bibitem[\protect\astroncite{{Miszalski} et~al.}{2009a}]{Miszalski:2009eu}
{Miszalski}, B., {Acker}, A., {Moffat}, A.~F.~J., {Parker}, Q.~A., and
  {Udalski}, A.: 2009a,
\newblock {\em \aap} {\bf 496}, 813

\bibitem[\protect\astroncite{{Miszalski} et~al.}{2009b}]{Miszalski:2009oq}
{Miszalski}, B., {Acker}, A., {Parker}, Q.~A., and {Moffat}, A.~F.~J.: 2009b,
\newblock {\em ArXiv e-prints}

\bibitem[\protect\astroncite{{Moe} and {De Marco}}{2006}]{Moe:2006fc}
{Moe}, M. and {De Marco}, O.: 2006,
\newblock {\em \apj} {\bf 650}, 916

\bibitem[\protect\astroncite{{Mullally} et~al.}{2009}]{Mullally:2009uq}
{Mullally}, F., {Reach}, W.~T., {De Gennaro}, S., and {Burrows}, A.: 2009,
\newblock {\em \apj} {\bf 694}, 327

\bibitem[\protect\astroncite{{Mullally} et~al.}{2008}]{Mullally:2008fk}
{Mullally}, F., {Winget}, D.~E., {De Gennaro}, S., {Jeffery}, E., {Thompson},
  S.~E., {Chandler}, D., and {Kepler}, S.~O.: 2008,
\newblock {\em \apj} {\bf 676}, 573

\bibitem[\protect\astroncite{{Niedzielski} et~al.}{2009}]{Niedzielski:2009lr}
{Niedzielski}, A., {Go{\'z}dziewski}, K., {Wolszczan}, A., {Konacki}, M.,
  {Nowak}, G., and {Zieli{\'n}ski}, P.: 2009,
\newblock {\em \apj} {\bf 693}, 276

\bibitem[\protect\astroncite{{Nordhaus} and {Blackman}}{2006}]{Nordhaus:2006oq}
{Nordhaus}, J. and {Blackman}, E.~G.: 2006,
\newblock {\em \mnras} {\bf 370}, 2004

\bibitem[\protect\astroncite{{Nordhaus} et~al.}{2007}]{Nordhaus:2007il}
{Nordhaus}, J., {Blackman}, E.~G., and {Frank}, A.: 2007,
\newblock {\em \mnras} {\bf 376}, 599

\bibitem[\protect\astroncite{{Nordhaus} et~al.}{2008a}]{Nordhaus:2008fe}
{Nordhaus}, J., {Busso}, M., {Wasserburg}, G.~J., {Blackman}, E.~G., and
  {Palmerini}, S.: 2008a,
\newblock {\em \apjl} {\bf 684}, L29

\bibitem[\protect\astroncite{{Nordhaus} et~al.}{2008b}]{Nordhaus:2008lr}
{Nordhaus}, J., {Minchev}, I., {Sargent}, B., {Forrest}, W., {Blackman}, E.~G.,
  {de Marco}, O., {Kastner}, J., {Balick}, B., and {Frank}, A.: 2008b,
\newblock {\em \mnras} {\bf 388}, 716

\bibitem[\protect\astroncite{{Ogilvie} and {Lin}}{2004}]{Ogilvie:2004sy}
{Ogilvie}, G.~I. and {Lin}, D.~N.~C.: 2004,
\newblock {\em \apj} {\bf 610}, 477

\bibitem[\protect\astroncite{{Ogilvie} and {Lin}}{2007}]{Ogilvie:2007kb}
{Ogilvie}, G.~I. and {Lin}, D.~N.~C.: 2007,
\newblock {\em \apj} {\bf 661}, 1180

\bibitem[\protect\astroncite{{Paczynski}}{1976}]{Paczynski:1976fj}
{Paczynski}, B.: 1976,
\newblock in {P.~Eggleton, S.~Mitton, \& J.~Whelan} (ed.), {\em Structure and
  Evolution of Close Binary Systems}, Vol.~73 of {\em IAU Symposium}, pp 75--+

\bibitem[\protect\astroncite{{Paxton}}{2004}]{Paxton:2004wd}
{Paxton}, B.: 2004,
\newblock {\em \pasp} {\bf 116}, 699

\bibitem[\protect\astroncite{{Peimbert}}{1990}]{Peimbert:1990ta}
{Peimbert}, M.: 1990,
\newblock {\em Revista Mexicana de Astronomia y Astrofisica} {\bf 20}, 119

\bibitem[\protect\astroncite{{Qian} et~al.}{2010}]{Qian:2010uq}
{Qian}, S., {Liao}, W., {Zhu}, L., {Dai}, Z., {Liu}, L., {He}, J., {Zhao}, E.,
  and {Li}, L.: 2010,
\newblock {\em \mnras} {\bf 401}, L34

\bibitem[\protect\astroncite{{Rasio} et~al.}{1996}]{Rasio:1996yq}
{Rasio}, F.~A., {Tout}, C.~A., {Lubow}, S.~H., and {Livio}, M.: 1996,
\newblock {\em \apj} {\bf 470}, 1187

\bibitem[\protect\astroncite{{Reimers}}{1975}]{Reimers:1975lr}
{Reimers}, D.: 1975,
\newblock {\em Memoires of the Societe Royale des Sciences de Liege} {\bf 8},
  369

\bibitem[\protect\astroncite{{Reyes-Ruiz} and
  {L{\'o}pez}}{1999}]{Reyes-Ruiz:1999lr}
{Reyes-Ruiz}, M. and {L{\'o}pez}, J.~A.: 1999,
\newblock {\em \apj} {\bf 524}, 952

\bibitem[\protect\astroncite{{Rybicki} and {Denis}}{2001}]{Rybicki:2001fk}
{Rybicki}, K.~R. and {Denis}, C.: 2001,
\newblock {\em Icarus} {\bf 151}, 130

\bibitem[\protect\astroncite{{Sabin} et~al.}{2007}]{Sabin:2007zp}
{Sabin}, L., {Zijlstra}, A.~A., and {Greaves}, J.~S.: 2007,
\newblock {\em \mnras} {\bf 376}, 378

\bibitem[\protect\astroncite{{Sackmann} et~al.}{1993}]{Sackmann:1993lr}
{Sackmann}, I., {Boothroyd}, A.~I., and {Kraemer}, K.~E.: 1993,
\newblock {\em \apj} {\bf 418}, 457

\bibitem[\protect\astroncite{{Sahai} and {Trauger}}{1998}]{Sahai:1998ee}
{Sahai}, R. and {Trauger}, J.~T.: 1998,
\newblock {\em \aj} {\bf 116}, 1357

\bibitem[\protect\astroncite{{Sato} et~al.}{2008a}]{Sato:2008uq}
{Sato}, B., {Izumiura}, H., {Toyota}, E., {Kambe}, E., {Ikoma}, M., {Omiya},
  M., {Masuda}, S., {Takeda}, Y., {Murata}, D., {Itoh}, Y., {Ando}, H.,
  {Yoshida}, M., {Kokubo}, E., and {Ida}, S.: 2008a,
\newblock {\em \pasj} {\bf 60}, 539

\bibitem[\protect\astroncite{{Sato} et~al.}{2008b}]{Sato:2008qy}
{Sato}, B., {Toyota}, E., {Omiya}, M., {Izumiura}, H., {Kambe}, E., {Masuda},
  S., {Takeda}, Y., {Itoh}, Y., {Ando}, H., {Yoshida}, M., {Kokubo}, E., and
  {Ida}, S.: 2008b,
\newblock {\em \pasj} {\bf 60}, 1317

\bibitem[\protect\astroncite{{Schr{\"o}der} and
  {Cuntz}}{2005}]{Schroder:2005uq}
{Schr{\"o}der}, K. and {Cuntz}, M.: 2005,
\newblock {\em \apjl} {\bf 630}, L73

\bibitem[\protect\astroncite{{Schr{\"o}der} and
  {Cuntz}}{2007}]{Schroder:2007qy}
{Schr{\"o}der}, K. and {Cuntz}, M.: 2007,
\newblock {\em \aap} {\bf 465}, 593

\bibitem[\protect\astroncite{{Silvotti} et~al.}{2007}]{Silvotti:2007fk}
{Silvotti}, R., {Schuh}, S., {Janulis}, R., {Solheim}, J., {Bernabei}, S.,
  {{\O}stensen}, R., {Oswalt}, T.~D., {Bruni}, I., {Gualandi}, R., {Bonanno},
  A., {Vauclair}, G., {Reed}, M., {Chen}, C., {Leibowitz}, E., {Paparo}, M.,
  {Baran}, A., {Charpinet}, S., {Dolez}, N., {Kawaler}, S., {Kurtz}, D.,
  {Moskalik}, P., {Riddle}, R., and {Zola}, S.: 2007,
\newblock {\em \nat} {\bf 449}, 189

\bibitem[\protect\astroncite{{Soker}}{1995}]{Soker:1995lr}
{Soker}, N.: 1995,
\newblock {\em \mnras} {\bf 274}, 147

\bibitem[\protect\astroncite{{Soker}}{1996}]{Soker:1996fk}
{Soker}, N.: 1996,
\newblock {\em \apjl} {\bf 460}, L53+

\bibitem[\protect\astroncite{{Spiegel} et~al.}{2010}]{Spiegel:aa}
{Spiegel}, D.~S., {Goodman}, J., {Ibgui}, L., {Nordhaus}, J., and {Burrows},
  A.: 2010,
\newblock {\em In prep.}

\bibitem[\protect\astroncite{{Tremblay} and {Bergeron}}{2007}]{Tremblay:2007fr}
{Tremblay}, P. and {Bergeron}, P.: 2007,
\newblock {\em \apj} {\bf 657}, 1013

\bibitem[\protect\astroncite{{van Winckel}}{2003}]{van-Winckel:2003pi}
{van Winckel}, H.: 2003,
\newblock {\em \araa} {\bf 41}, 391

\bibitem[\protect\astroncite{{Verbunt} and {Phinney}}{1995}]{Verbunt:1995rt}
{Verbunt}, F. and {Phinney}, E.~S.: 1995,
\newblock {\em \aap} {\bf 296}, 709

\bibitem[\protect\astroncite{{Villaver} and {Livio}}{2007}]{Villaver:2007lr}
{Villaver}, E. and {Livio}, M.: 2007,
\newblock {\em \apj} {\bf 661}, 1192

\bibitem[\protect\astroncite{{Villaver} and {Livio}}{2009}]{Villaver:2009qy}
{Villaver}, E. and {Livio}, M.: 2009,
\newblock {\em \apjl} {\bf 705}, L81

\bibitem[\protect\astroncite{{Vlemmings} et~al.}{2006}]{Vlemmings:2006vl}
{Vlemmings}, W.~H.~T., {Diamond}, P.~J., and {Imai}, H.: 2006,
\newblock {\em \nat} {\bf 440}, 58

\bibitem[\protect\astroncite{{Vlemmings} and {van
  Langevelde}}{2008}]{Vlemmings:2008ad}
{Vlemmings}, W.~H.~T. and {van Langevelde}, H.~J.: 2008,
\newblock {\em \aap} {\bf 488}, 619

\bibitem[\protect\astroncite{{Witt} et~al.}{2009}]{Witt:2009wd}
{Witt}, A.~N., {Vijh}, U.~P., {Hobbs}, L.~M., {Aufdenberg}, J.~P., {Thorburn},
  J.~A., and {York}, D.~G.: 2009,
\newblock {\em \apj} {\bf 693}, 1946

\bibitem[\protect\astroncite{{Yoder} and {Peale}}{1981}]{Yoder:1981zv}
{Yoder}, C.~F. and {Peale}, S.~J.: 1981,
\newblock {\em Icarus} {\bf 47}, 1

\bibitem[\protect\astroncite{{Zahn}}{1989}]{Zahn:1989lr}
{Zahn}, J.: 1989,
\newblock {\em \aap} {\bf 220}, 112

\bibitem[\protect\astroncite{{Zahn}}{1966}]{Zahn:1966jk}
{Zahn}, J.~P.: 1966,
\newblock {\em Annales d'Astrophysique} {\bf 29}, 489

\bibitem[\protect\astroncite{{Zapolsky} and {Salpeter}}{1969}]{Zapolsky:1969hs}
{Zapolsky}, H.~S. and {Salpeter}, E.~E.: 1969,
\newblock {\em \apj} {\bf 158}, 809

\end{thebibliography}
\bibliographystyle{astron}

\end{document}